\def\notes{1}

\documentclass[sigconf,nonacm,authorversion,balance=false]{acmart}

\makeatletter
\def\@ACM@checkaffil{
    \if@ACM@instpresent\else
    \ClassWarningNoLine{\@classname}{No institution present for an affiliation}%
    \fi
    \if@ACM@citypresent\else
    \ClassWarningNoLine{\@classname}{No city present for an affiliation}%
    \fi
    \if@ACM@countrypresent\else
        \ClassWarningNoLine{\@classname}{No country present for an affiliation}%
    \fi
}
\makeatother

\usepackage{mdframed} 
\usepackage{amsfonts}
\usepackage{amsmath}
\usepackage{eucal}
\usepackage{lipsum}
\usepackage{enumitem}
\usepackage[normalem]{ulem}
\usepackage{dblfloatfix}

\usepackage{xspace}
\usepackage{booktabs}
\usepackage{multirow}
\usepackage{algorithm}
\usepackage{algpseudocode}
\usepackage{hyperref}
\usepackage{cleveref} 
\usepackage{amsthm}
\usepackage{graphicx}
\graphicspath{ {./images/} }
\usepackage{tikz}
\usepackage{tikzpeople}
\usepackage{pgffor}
\usepackage{pgfmath}
\usepackage{pgfplots}
\usepackage{subfigure}
\usepackage[Tol,OkabeIto,pgf]{colorblind}

\usepackage{ifxetex}
\ifxetex
  \usepackage{fontspec}
  \usepackage{xunicode}
  \usepackage{xltxtra}
\else
  \usepackage[utf8]{inputenc}
  \usepackage[T1]{fontenc}
\fi

\newcommand{\censor}[1]{[redacted for anonymous submission]\xspace}

\pgfplotsset{width=7cm,compat=1.18}
\usepgfplotslibrary{dateplot}
\usepgfplotslibrary{fillbetween}

\usetikzlibrary{positioning}
\usetikzlibrary{shadows}
\usetikzlibrary{shapes.geometric}
\usetikzlibrary{shapes.symbols}
\usetikzlibrary{shapes.arrows}
\usepackage[tableposition=top]{caption}

\input{tikz/database} 
\input{tikz/smartphone} 

\tikzstyle{MechanismBoxArch} = [rectangle, draw, minimum width=2.8cm, minimum height=1cm, text width=2.7cm, align=center]
\tikzstyle{MechanismBox} = [rectangle, draw, minimum width=2cm, minimum height=1cm, text width=6em, align=center]
\tikzstyle{QueryBox} = [rectangle, rounded corners, draw, minimum width=1.5cm, minimum height=0.75cm]

\def\ColorGaussPerturb{T-Q-MC7} 
\def\ColorLapNoise{T-Q-MC6} 
\def\ColorQuery{T-Q-MC5} 

\def\ColorGraphPlaintext{T-Q-MC0}

\def\ColorGraphFC{\ColorLapNoise}
\def\ColorGraphFine{\ColorLapNoise}
\def\ColorGraphCC{\ColorGaussPerturb}
\def\ColorGraphCoarse{\ColorGaussPerturb}

\def\ColorPerturb{\ColorGaussPerturb}
\def\ColorPerturbed{\ColorGaussPerturb}

\theoremstyle{definition}
\newtheorem{definition}{Definition}[section]

\crefname{definition}{Def.}{Defs.}
\crefname{figure}{Fig.}{Figs.}
\crefname{section}{Sec.}{Secs.}
\crefname{equation}{Eq.}{Eqs.}
\crefname{appendix}{App.}{Apps.}

\definecolor{mypurple}{rgb}{0.58,0.23,0.94}
\ifnum\notes=1
\newcommand{\mx}[1]{\textcolor{red}{[MX: #1]}}
\newcommand{\pj}[1]{\textcolor{mypurple}{[PJ: #1]}}
\newcommand{\mg}[1]{\textcolor{cyan}{[MG: #1]}}
\newcommand{\sstodo}[1]{\textcolor{blue}{[SS: #1]}}
\newcommand{\mgtodo}[1]{\textcolor{green}{[MG: #1]}}
\else
\newcommand{\mx}[1]{ }
\newcommand{\pj}[1]{ }
\newcommand{\mg}[1]{ }
\newcommand{\sstodo}[1]{ }
\newcommand{\mgtodo}[1]{ }
\fi

\newcommand{\ignore}[1]{}

\newcommand{\rot}[1]{\rotatebox{35}{#1}}

\newcommand{\CountFine}{\textbf{Query-Count-Fine}\xspace}
\newcommand{\CountCoarse}{\textbf{Query-Count-Coarse}\xspace}
\newcommand{\ThresholdFine}{\textbf{Query-Threshold-Fine}\xspace}
\newcommand{\ThresholdCoarse}{\textbf{Query-Threshold-Coarse}\xspace}

\newcommand{\QueryFineCount}{\CountFine}
\newcommand{\QueryCoarseCount}{\CountCoarse}
\newcommand{\QueryFineThreshold}{\ThresholdFine}
\newcommand{\QueryCoarseThreshold}{\ThresholdCoarse}

\newcommand{\BallRadius}{\ensuremath{a}\xspace}

\newcommand{\Query}{\ensuremath{q}\xspace}

\newcommand{\JLTk}{\ensuremath{k}\xspace}
\newcommand{\EmbDimension}{\ensuremath{k}\xspace}

\newcommand{\EmbDimFinal}{\EmbDimension}
\newcommand{\EmbDimInitial}{\ensuremath{\ell}\xspace}
\newcommand{\EmbLength}{\ensuremath{\ell}\xspace}
\newcommand{\EmbLen}{\EmbLength}
\newcommand{\NumMsgs}{\ensuremath{N}\xspace}

\newcommand{\Projection}{\ensuremath{P}\xspace}

\newcommand{\NoiseMatrix}{\ensuremath{\Delta}\xspace}

\newcommand{\eps}{\epsilon\xspace}
\newcommand{\ScalingFactor}{\ensuremath{b}\xspace}
\newcommand{\Lap}{\ensuremath{\text{Lap}}\xspace}
\newcommand{\Dist}{\ensuremath{\textsf{dist}}\xspace}

\newcommand{\Epoch}{e\xspace}
\newcommand{\EpsilonCoarse}{\eps_P\xspace}
\newcommand{\DeltaCoarse}{\delta_P\xspace}
\newcommand{\EpsilonCountCoarse}{\EpsilonCoarse\xspace}
\newcommand{\EpsilonCountFine}{\eps_C\xspace}
\newcommand{\EpsilonFineCount}{\EpsilonCountFine\xspace}
\newcommand{\EpsilonFineCounts}{\EpsilonCountFine\xspace}
\newcommand{\EpsilonFineThreshold}{\eps_T\xspace}
\newcommand{\EpsilonCoarseTotal}{\EpsilonCoarse\xspace}
\newcommand{\EpsilonFineTotal}{\eps_{F}\xspace}
\newcommand{\EpsilonFEpoch}{\eps_{F,\Epoch}\xspace}

\newcommand{\DBFineEpoch}{D_{F,\Epoch}\xspace}
\newcommand{\DBCoarseEpoch}{D_{P,\Epoch}\xspace}

\newcommand{\JLParam}{\alpha_{\textsf{JL}}\xspace}

\newcommand{\NParties}{n\xspace}

\newcommand{\Synopsis}{Synopsis\xspace}
\newcommand{\SVS}{\Synopsis}
\newcommand{\SecretVectorSearch}{\Synopsis}

\newcommand{\cmark}{\checkmark}

\begin{document}

\title{Synopsis: Secure and private trend inference from encrypted semantic embeddings}

\author{Madelyne Xiao}
\affiliation{Princeton University}

\author{Palak Jain}
\affiliation{Boston University}

\author{Micha Gorelick}
\affiliation{Digital Witness Lab}

\author{Sarah Scheffler}
\affiliation{Carnegie Mellon University}

\maketitle

\hyphenation{Whats-App}
\hyphenation{Mech-anism}
\hyphenation{mech-anism}

\date{}

WhatsApp and many other commonly used communication platforms guarantee end-to-end encryption (E2EE), which requires that service providers lack the cryptographic keys to read communications on their own platforms. WhatsApp's privacy-preserving design makes it difficult to study important phenomena like the spread of misinformation or political messaging, as users have a clear expectation and desire for privacy and little incentive to forfeit that privacy in the process of handing over raw data to researchers, journalists, or other parties.

We introduce \textit{\Synopsis}, a secure architecture for analyzing messaging trends in consensually-donated E2EE messages using message embeddings. 
Since the goal of this system is investigative journalism workflows, \Synopsis must facilitate both exploratory and targeted analyses---a challenge for systems using differential privacy (DP)~\cite{cummings2024advancing, nanayakkara2024measure}, and, for different reasons, a challenge for private computation approaches based on cryptography. To meet these challenges, we combine techniques from the local and central DP models and wrap the system in malicious-secure multi-party computation to ensure the DP query architecture is the only way to access messages, preventing any party from directly viewing stored message embeddings.

Evaluations on a dataset of Hindi-language WhatsApp messages (34,024 messages represented as 500-dimensional embeddings) demonstrate the efficiency and accuracy of our approach. Queries on this data run in about 30 seconds, and the accuracy of the fine-grained interface exceeds 94\% on benchmark tasks.

\section{Introduction}
\label{sec:intro}

WhatsApp is the dominant messaging app in the majority world, with nearly three billion monthly active users in 2024 and reported market share in 2022 of 98.9\% in Brazil and 97.1\% in India \cite{statistaWhatsAppMarket}.
Given its prevalence, journalists and researchers frequently seek to analyze messaging behavior on WhatsApp, especially during politically charged times like elections and other important events.

It is difficult, even in ideal circumstances, for researchers to gain access to social media messaging data, but it is especially difficult in the case of WhatsApp due to WhatsApp's use of end-to-end encryption (E2EE) \cite{wa_about_e2ee,double_ratchet}. In E2EE, the messaging service provider does not possess the cryptographic keys to read message contents; these are held solely by the ``end'' users of the communication.

The privacy of this communication makes it both more difficult and more important to identify trends in these communications.
In many recent elections, WhatsApp communications played a major role in both mainstream campaigning and also in the spread of misinformation \cite{domingo2023india,shih2023inside,lamb2024generative,sekargati2024hate}.
Researchers have surfaced two major angles by which analysis of WhatsApp activity is possible.

The first approach to researching WhatsApp activity (both for scientific and journalistic purposes) is using publicly accessible data or group chats with public invitation URLs \cite{bursztynThousandsOfSmall,de2021digital,agarwalJettisoningJunk}, under the assumption that these public groups have minimal privacy expectations.
However, there is reason to believe that behavior in private groups differs significantly from that in public groups, where observer effects and self-censorship likely played a non-trivial role.
Since the majority of WhatsApp messages are sent as direct messages or in small private groups rather than in public groups \cite{rosenfeld2018studywhatsappusagepatterns,seufert2015analysis}, analysis of WhatsApp activity that is limited to public groups is likely to be non-representative of the full spectrum of on-platform activity, and excludes private communications.

A second possible solution to issues of access and observation of WhatsApp activity is \emph{data donation}. 
While some donations are limited to individual messages and function more like a tipline \cite{kazemi2022a,arun2019whatsapp}, some newer approaches to data donation are longer-term.
Under the terms of consensual donation agreements brokered by researchers and data stewards, donors make an informed decision to provide researchers with ongoing automated access to some or all of their messaging data.
This sort of data has already proven useful in studies about platforms and political speech online \cite{themarkupBuiltFacebook,garimella2024whatsappexplorerdatadonation,whatsapp_monitor}.

However, data donation also poses privacy concerns.
A naive approach to data donation simply gives the researcher full access to all messages the donor has sent or received on WhatsApp.
This text messaging data is high-sensitivity for several reasons:
First, the dataset includes text messages that may have been donated by the \emph{receiver} of a message without the message \emph{sender's} consent.
Second, the text messages were not previously available to any third parties including the service provider due to end-to-end encryption, so their analysis by a researcher reflects a significant change from the status quo.
Third, the dataset is likely to include a significant amount of information in miscellaneous text messages that is not relevant to any particular journalistic investigation.
And fourth, the dataset will likely contain sensitive political discussions of the messages involved, and may carry legal risk to both the senders/receivers of the messages, the data donors, and the researchers, depending on the local laws regarding political content.

This raises the following question: \emph{What is the right way to provide privacy to donated E2EE messages while still enabling research?}

In the naive approach to data donation, the data donor (an ``end'' of the end-to-end encryption) relays all messages she receives to a researcher for analysis.
The researcher retains the raw or lightly-redacted message data, analyzes it, and possibly releases partially-redacted versions of the dataset to others.
Indeed, this approach is used by some researchers researching WhatsApp texts \cite{whatsapp_monitor,garimella2024whatsappexplorerdatadonation} and internet browsing behavior \cite{nio,feal2024introduction}.

However, privacy protections for full donated messages \emph{against researchers or journalists} are thin, and messages can be used for purposes other than aggregate analysis. The redaction methods of these systems  use pattern-based detection to remove phone numbers, addresses, or names \cite{whatsapp_monitor,google_cdlp}. However, this ``sanitization'' approach is naturally brittle to non-standard ways of describing sensitive information. More importantly, pattern-based redaction by definition does not apply to non-pattern-based notions of sensitive information (see e.g. \cite{dwork_2006,ohm2009broken}), among other issues.
Simple AI tools or reliance on converting messages to ``embedding'' vectors are also not sufficient to protect privacy \cite{song2020information}; these embedding vectors are often reversible to the exact message itself \cite{morris2023text,morris2023language,hitaj_deep_2017,li_towards_2018}.
However, these embedding can be helpful with analysis as embedding models can be selected to help with multi-lingual datasets, multi-modal datasets or with datasets with colloquial or short-form writing.
A different approach is required for processing messages in aggregate.

Our system, \Synopsis, accomplishes this goal by enabling journalists to look for \textit{trends in message data without ever accessing individual texts} or being in a position to leak information about them.  This also gives platform users the ability to donate their data for the public good without exposing the contents of their messages. Journalists are able to write natural language queries and see counts of how many semantically similar messages, or whether that count has reached a threshold, are present in the donated corpus.
In addition to its benefits for private group analysis, our aggregate privacy-preserving approach also provides an alternative to direct publication of scraped data from public groups, especially when there are concerns about the contents of those channels and the presence or absence of informed consent from group participants. \Synopsis provides stronger privacy protections for individual message senders than does straightforward anonymization of account names and other identifiers---this anonymization approach was recently adopted by researchers who released a corpus of 2 billion Discord messages in plaintext, including many messages from underage users \cite{gault_2025, aquino2025discord}.

As shown in \Cref{fig:architecture}, \Synopsis enables four types of queries to text corpora, each conferring different degrees of privacy and accuracy. These query types are based on three differentially-private (DP) mechanisms and are implemented under cryptographic protection of malicious-secure multi-party computation.

Synopsis's ``fine-grained'' query regime in the central DP model allows researchers to issue targeted queries to an unperturbed and encrypted dataset; resulting counts of matching elements are noised before release, burning ``privacy budget.'' A complementary ``coarse-grained'' query regime in the local DP model allows researchers to issue an unbounded number of queries to a perturbed and encrypted dataset of message embeddings. This regime is ideal for consistent observation and exploratory data analysis, but is less accurate overall and tends to overestimate results (see \Cref{sec:cc-overestimate}). Taken together, these regimes permit queriers to flexibly toggle between exploratory and targeted analysis modes while preserving the privacy of message contents and senders.

Synopsis is one part of a larger system of investigative data journalism tooling, called WhatsAppWatch. The tool is already in place to allow researchers and journalists to monitor and query data donations from public WhatsApp groups. Data usage agreements for the public data used for this study were conducted within the structure of this existing tooling.

\subsection{Our contributions}

In this work, we design and implement \Synopsis, a privacy-preserving system that permits trend-based analysis of text message data received from clients of an end-to-end encrypted messaging application. 
Our technical contributions are as follows:
\begin{itemize}
    \item \textbf{Methodology for designing our privacy-conscious architecture for the under-examined use case of data journalism.}
    In \Cref{sec:methodology}, we lay out our methodology for deciding on the appropriate architecture for investigating trends in corpora of E2EE messages, including differential privacy (DP) over messages and cryptographic protections via malicious-secure multi-party computation (MPC). We conduct a stakeholder analysis and discuss the subtle design considerations resulting from this analysis.
    We further discuss why we believe the \textit{consensual data donation} model is an appropriate choice for E2EE message analysis, why it should be considered \textit{best practice to provide technical privacy protections} to those consensual donations, \textit{which technical protections} to provide, and \textit{how to balance this with the need for useful outputs}.
    
    \item \textbf{Secure architecture design for enabling privacy-preser-ving semantic trend analysis.}
    We provide a flexible query algorithm utilizing a combination of central and local differential privacy (DP), and malicious-secure multi-party computation (MPC), as described in  In \Cref{sec:architecture}. We provide four query types:
    \begin{itemize}
    \item Fine-grained count queries (FC) for high-accuracy central-DP trend data, spending $\EpsilonFineCount$ privacy budget per query.
    \item Fine-grained threshold queries (FT) for high-accuracy central-DP queries. These queries return a Boolean response: ``1'' if the count of matching messages exceeds a threshold, and ``0'' otherwise. FT queries spend $\EpsilonFineThreshold$ privacy budget \emph{only} if the query returns 1.
    \item Coarse-grained count queries (CC) for low-accuracy local-DP trend data; no budget spend per query.
    \item Coarse-grained threshold queries (CT) for low-accuracy local-DP threshold queries; no budget spend per query.
    \end{itemize}
    In all cases, MPC ensures that the only output to the system is the query result (typically a trend line measuring counts of donated messages that match a query per epoch---see \Cref{fig:trendlines} for an example).
    This design permits both exploratory and targeted trend analysis while providing privacy and security to the message database and donors.

    \item \textbf{Implementation, usage scenario, and deployment.}
    In \Cref{sec:implementation} we discuss implementation
    and provide a demonstration of \Synopsis through testing on a real world corpus of donated WhatsApp messages from public chats.
    Our code is available and open-source.\footnote{Prototype code available here: 
    \url{https://github.com/digital-witness-lab/synopsis} 
    } Trend lines from an exemplar investigation are shown in\  \Cref{fig:trendlines}. 
\end{itemize}

\newcommand{\monthname}[1]{\ifcase#1\or Jan\or Feb\or Mar\or Apr\or May\or Jun\or Jul\or Aug\or Sep\or Oct\or Nov\or Dec\fi}
\newcommand{\shortyear}[1]{\ifnum#1=2023 23\else\ifnum#1=2024 24\fi\fi}

\begin{figure*}[!htb]
    \centering
    \setlength{\parindent}{0ex}

    \begin{tikzpicture} 
        \begin{axis}[
            scale only axis, 
            height=3.5cm,
            width=\textwidth*0.92,
            grid=both,
            max space between ticks=40, 
            xlabel = {Date (MM-YY)},
            ylabel = {Count of messages matching query},
            xmin = 2023-06-20,
            xmax = 2024-02-20,
            minor x tick num=1,
            minor y tick num=4,
            major tick length=0.15cm,
            minor tick length=0.075cm,
            tick style={semithick,color=black}, 
            date coordinates in=x,
            legend style={at={(0.01, 0.78)},anchor=west, font=\footnotesize},
            xticklabel=\month-\shortyear{\year}
            ]

        \addplot[very thick, color = \ColorGraphPlaintext] 
        table[x=day,y=count] {plotdata/plaintext_counts.dat};
        \addlegendentry{Ground truth query matches}
        
        \addplot[
        line cap=round, color = \ColorGraphFine] 
        table[x=day,y=row_mean] {plotdata/fc_ten_trials.dat};
        \addlegendentry{FC query matches $\EpsilonFineCounts = 0.6$}
        \addplot [name path=upper,draw=none, forget plot] table[x=day,y expr=\thisrow{row_mean}+\thisrow{row_std}] {plotdata/fc_ten_trials.dat};
        \addplot [name path=lower,draw=none, forget plot] table[x=day,y expr=\thisrow{row_mean}-\thisrow{row_std}] {plotdata/fc_ten_trials.dat};
        \addplot [fill=\ColorGraphFine!20, fill opacity=0.5, forget plot] fill between[of=upper and lower];

        \addplot[dotted, dash pattern=on 2.3pt off 1pt, line cap=round, color = \ColorGraphCoarse] 
        table[x=day,y=row_mean] {plotdata/cc_ten_trials.dat};
        \addlegendentry{CC query matches (7-day centered average), $\EpsilonCoarseTotal = 2$}
        \addplot [name path=upper,draw=none, forget plot] table[x=day,y expr=\thisrow{row_mean}+\thisrow{row_std}] {plotdata/cc_ten_trials.dat};
        \addplot [name path=lower,draw=none, forget plot] table[x=day,y expr=\thisrow{row_mean}-\thisrow{row_std}] {plotdata/cc_ten_trials.dat};
        \addplot [fill=\ColorGraphCoarse!20, fill opacity=0.5, forget plot] fill between[of=upper and lower];
        
        \end{axis}
        \end{tikzpicture}
\vspace{-0.5em}
\caption{Comparison of fine-grained and coarse-grained analyses of Ram temple-related messaging rates from June 2023 to February 2024. Fine-grained and plaintext trendlines show daily counts; the coarse-grained trendline shows a centered 7-day rolling average. We discuss the reasoning for the CC queries' overestimate of the FC queries---the effect of ``surrounding'' vectors that are nearby but not within the query radius---in \Cref{sec:cc-overestimate}.}
\label{fig:trendlines}
\end{figure*}
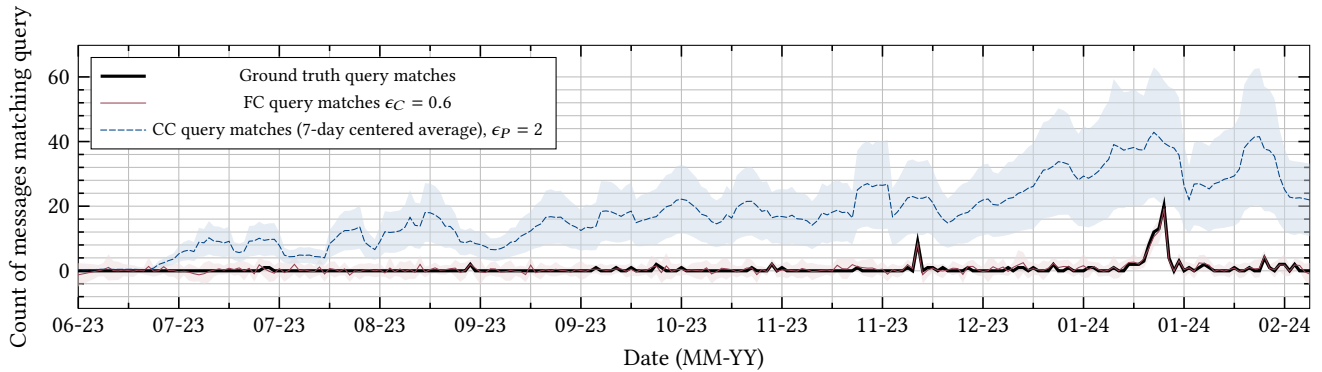
\section{Related Work}
\label{sec:related-work}

This work draws on three existing lines of research: (1) investigative journalism involving end-to-end encryption, (2) private analysis of text corpora and device data under differential privacy, and (3) cryptographic protections for data analysis in which the data must remain confidential to both an external querier and the server storing the data.

\subsection{Investigative Journalism in WhatsApp}

WhatsApp has been the subject of significant investigative journalism efforts, in part because of instances in which inflammatory, misinformative, or illicit online speech led to serious consequences offline. For example, in India, doctored or fake videos of the Ram Temple inauguration (discussed further in~\Cref{sec:investigation}) and ads for illegal firearms reached millions of users through public and private chats \cite{shekhar_guns_2024, ram_scam_2024}.
Prospective and ongoing measurement of public WhatsApp groups has demonstrated the scope of political propaganda leading up to elections in Brazil and India from official and ``surrogate'' sources as measured in both official and unofficial public groups \cite{avelar_bolsonaro_2019, jaswal_bjp_2024,mattu_bjp_2024}.
In another form of investigation that is more retrospective, a journalistic investigation of the spread of specific misleading videos that led to the murder of journalist Gauri Lankesh happened years after her actual death, when video documentation of her death surfaced \cite{lankesh1,lankesh2}.

For both exploratory and retrospective use cases, compared to public groups, it is significantly harder to measure the spread of this messaging in private groups, which are responsible for the overwhelming majority of WhatsApp messages sent \cite{rosenfeld2018studywhatsappusagepatterns,seufert2015analysis}.
One method that has emerged to address these questions is \emph{data donation} (e.g. \cite{whatsapp_monitor,garimella2024whatsappexplorerdatadonation}). 
In the data donation-based WhatsApp Explorer application \cite{garimella2024whatsappexplorerdatadonation} (formerly WhatsApp Monitor \cite{whatsapp_monitor}), one central server maintains a database of complete full messages; the message sender receives a pseudonym; and full message contents are retained, with the exception of email addresses and phone numbers, which are redacted via pattern-matched  detection in the Google Cloud Data Loss Protection API \cite{google_cdlp}.
This approach is similar to the approach taken by the National Internet Observatory, which measures Internet behavior on an ongoing basis using a browser extension \cite{nio,feal2024introduction}.

\subsection{Private text analysis}

Early work on constructing private text corpora with limited query capability focused on the problem of removing personally identifiable information from medical records accessible to hospital administrators or external researchers \cite{sweeney1996replacing,douglass2005identification}.
From these early works, researchers and practitioners developed a variety of heuristic notions of text privacy that are still in use today, including rule-based or AI-based text sanitation \cite{douglass2005identification,deleger2013large,dernoncourt2017identification,johnson2020deidentification,anandan2012t,chakaravarthy2008efficient,sanchez2016c}.

Differential privacy (DP) \cite{dwork_2006} drastically reformed many fields of privacy, including text analysis. As defined in \Cref{sec:preliminaries}, DP guarantees that running an algorithm on datasets that are ``neighboring'' (i.e., differing in only one input; see \Cref{def:dp} and \Cref{sec:unit-of-privacy}) does not change the output of the algorithm too much.  
The extent of this protection is parameterized by $\epsilon$ and $\delta$ as defined in \Cref{def:dp}.

Although much work in DP text analysis treats the notion of ``neighboring'' datasets as those that differ by a single word
(e.g., \cite{feyisetan_perturb_2020,xu_density-aware_2021,xu_utilitarian_2021}), token (e.g., \cite{yue_differential_2021,qu_natural_2021,chen_customized_2023}), or sentence (e.g., \cite{meehan_sentence-level_2022,weggenmann_dp-vae_2022,krishna_adept_2021,habernal_when_2021}), these word, token, and sentence-level approaches still reveal a significant amount of information about individual messages \cite{hitaj_deep_2017,li_towards_2018,mattern_limits_2022}.
Additionally, word- and sentence-level approaches are less applicable to many modern text analysis methods, since modern language models are able to ``embed'' an entire message in a fixed-length, bounded embedding vector, where there are no one-to-one correspondences between lexical units (words, sentences) and embedding dimensions.
(We also restate that the embedding vector itself does not provide privacy and, in many cases, can be used to reconstruct the exact text used to generate it \cite{morris2023text,morris2023language,hitaj_deep_2017,li_towards_2018}.)
As we discuss in \Cref{sec:unit-of-privacy}, we take our definition of ``neighboring'' to be text corpora differing by one message.

DP also admits varying adversarial models, with the two most common being the following: the standard \emph{central} model, in which an external analyst makes queries to a central database (which itself is held by a trusted curator and therefore stores data in the clear; and the \emph{local} model \cite{DBLP:journals/siamcomp/KasiviswanathanLNRS11}, wherein DP guarantees also hold against the untrusted curator (typically because each data subjects perturb their inputs locally to provide DP \textit{before} adding the data to the database).

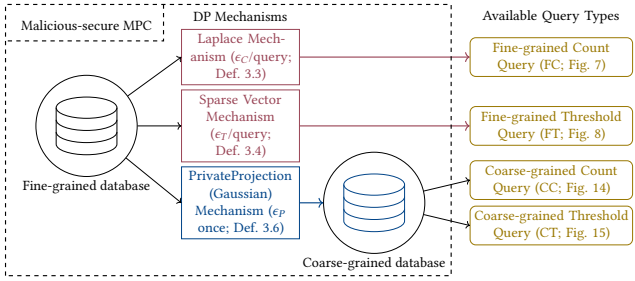
\begin{figure}[t!]
    \centering
    \resizebox{\linewidth}{!}{
    \begin{tikzpicture}[node distance=1.5cm, every node/.style={font=\large}]

\node[circle, draw, minimum size=2.5cm] (Circle1) {}; 
\node[database, scale=3] (Database1) at (Circle1) {};
\node[below of=Circle1] (Text1) {Fine-grained database};

Mechanism boxes
\node[MechanismBoxArch, right of=Circle1, xshift=2.3cm, yshift=0cm, color=\ColorLapNoise] (MechanismBoxSparse) {Sparse Vector Mechanism ($\EpsilonFineThreshold$/query; \Cref{def:svt})};
\node[MechanismBoxArch, above of=MechanismBoxSparse, yshift=0.2cm, color=\ColorLapNoise] (MechanismBoxLaplace) {Laplace Mechanism ($\EpsilonFineCount$/query; \Cref{def:laplace})};
\node[MechanismBoxArch,below of=MechanismBoxSparse, color=\ColorGaussPerturb, yshift=-0.4cm] (MechanismBoxGaussian) {PrivateProjection (Gaussian) Mechanism ($\EpsilonCoarseTotal$ once; \Cref{def:privateprojection})};
\node[above of=MechanismBoxLaplace, yshift=-0.5cm] (TextDP) {DP Mechanisms};

\node[circle, draw, minimum size=2.5cm, right of=MechanismBoxGaussian, xshift=1.8cm] (Circle2) {};
\node[database, scale=3, color=\ColorPerturbed] (Database2) at (Circle2) {};
\node[below of=Circle2] (Text2) {Coarse-grained database};

\def\QueryBoxWidth{3.8cm}
\node[QueryBox, above right of=Circle2, xshift=1.3cm, yshift=-0.5cm, color=\ColorQuery, text width=\QueryBoxWidth, anchor=west, align=center] (QueryCC) {Coarse-grained Count \\Query (CC; \Cref{alg:query-coarse-count})};
\node[QueryBox, below right of=Circle2, xshift=1.3cm, yshift=0.5cm, color=\ColorQuery, text width=\QueryBoxWidth, anchor=west, align=center] (QueryCT) {Coarse-grained Threshold\\Query (CT; \Cref{alg:query-coarse-threshold})};
\node[QueryBox, color=\ColorQuery, text width=\QueryBoxWidth, align=center] (QueryFC) at (MechanismBoxLaplace-|QueryCC) {Fine-grained Count\\Query (FC; \Cref{alg:query-fine-count})};
\node[QueryBox, color=\ColorQuery, text width=\QueryBoxWidth, align=center] (QueryFT) at (MechanismBoxSparse-|QueryCC) {Fine-grained Threshold\\Query (FT; \Cref{alg:query-fine-threshold})};
\node[above of=QueryFC, yshift=-0.5cm, align=center, text width=\QueryBoxWidth] (TextQuery) {Available Query Types};

\draw[->] (Circle1) -- (MechanismBoxLaplace.west);
\draw[->, color=\ColorLapNoise, thick] (MechanismBoxLaplace) -- (QueryFC.west);

\draw[->] (Circle1) -- (MechanismBoxSparse.west);
\draw[->, color=\ColorLapNoise, thick] (MechanismBoxSparse) -- (QueryFT.west);

\draw[->] (Circle1) -- (MechanismBoxGaussian.west);
\draw[->, color=\ColorPerturb, thick] (MechanismBoxGaussian) -- (Circle2);
\draw[->] (Circle2) -- (QueryCC.west);
\draw[->] (Circle2) -- (QueryCT.west);

\def\BoxTop{3}
\def\BoxBottom{-3.7}
\def\BoxLeft{-2}
\def\BoxRight{9}
\def\MPCRight{1.9}
\def\MPCBottom{2.1}
\draw[thick,dashed] (\BoxLeft,\BoxBottom) -- (\BoxLeft,\BoxTop) -- (\BoxRight,\BoxTop) -- (\BoxRight,\BoxBottom) -- cycle;
\node[xshift=2cm, yshift=-0.5cm, align=left] (MpcText) at (\BoxLeft,\BoxTop) {Malicious-secure MPC};
\draw[thick,dashed] (\MPCRight,\BoxTop) -- (\MPCRight, \MPCBottom) -- (\BoxLeft,\MPCBottom);

\end{tikzpicture}
    } 
    \caption{Our architecture: Four query types enabled by three differentially private mechanisms, all wrapped in malicious-secure multi-party computation between journalist queriers and MPC servers run by the data stewards. Technical details are found in \Cref{sec:architecture} and design choices in \Cref{sec:design}.}
    \label{fig:queries-to-mechanisms}
    \label{fig:query-to-mechanisms}
    \label{fig:queries-to-mechanism}
    \label{fig:mechanisms-to-queries}
    \label{fig:architecture}
    \vspace{-0.5em}
\end{figure}

\subsection{Cryptography for private data analysis}

In addition to DP, our analysis in \Cref{sec:design} led us to pursue cryptographic protections for donated E2EE texts beyond DP protections in order to ensure that the query interface is the \emph{only} way to access the text corpus, even for data stewards maintaining the database.

We ultimately implemented our own query architecture using malicious secure multi-party computation (in the MP-SPDZ library \cite{mp-spdz}) after rejecting significantly slower alternatives for generic private computation with similar security guarantees, e.g. homomorphic and functional encryption (see \Cref{sec:why-mpc}).

In \Cref{tab:relatedworks}, we chart the existing landscape of methods that permit queries over data that must remain private to the querier.
Synopsis is the first work to achieve all goals in \Cref{tab:relatedworks} simultaneously.
In particular, arbitrary retrospective queries would not be possible in works that only permit querying using a fixed set of keywords known in advance, or works that only enable tracking of the most frequently occurring messages.  
Secure sketching works and close relatives \cite{melis_efficient,hagen_p2kmv,boneh_lightweight,whisper} come close to the functionality we want: these approaches enable a querier to obtain aggregate private counts of particular items, which could be used over epochs to observe longitudinal trends. However, all of these works only allow queries for items known in advance (\cite{melis_efficient}) or for heavy hitters \cite{boneh_lightweight,whisper} rather than arbitrary retrospective queries.
Other applied cryptography works that address information retrieval (e.g. \cite{compass,tiptoe,DBLP:conf/sp/Servan-Schreiber22}) aim to solve a different problem: enabling a secret query over data held by a remote server, with the database viewable to the querier.  
Similar issues exist within a vast literature on searchable encryption that does not match our threat model (e.g. \cite{DBLP:conf/ndss/WuL23, DBLP:conf/osdi/DautermanFLPS20,DBLP:conf/sosp/AhmadSAAG21,DBLP:conf/sp/MishraPCCP18}); additionally, we prefer calculations on semantic embeddings to exact keyword matching on plaintext.

Prochlo \cite{bittau2017prochlo} is the closest to \Synopsis, as it also presents a combined cryptographic and differentially private system for performing data analysis, following an encode-shuffle-analyze paradigm; this paradigm could enable both prospective and retrospective queries.  However, it leaves the actual ``analysis'' generic, instead focusing on ``encoding and shuffling'' intake steps. \Synopsis answers the question: \textit{What versions of encoding and analysis are sufficiently useful and private for investigative journalism on WhatsApp?}

\begin{table}
\resizebox{0.8\linewidth}{!}{
\begin{tabular}{p{0.35\linewidth}p{0.05\linewidth}p{0.05\linewidth}p{0.05\linewidth}p{0.05\linewidth}p{0.05\linewidth}p{0.05\linewidth}}
\textbf{Work} & \rot{\textbf{Measures counts?}} &  \rot{\textbf{Retro. queries?}} & \rot{\textbf{Data priv. to querier?}} & \rot{\textbf{Non-exact matches?}} & \rot{\textbf{Malicious security?}} & \rot{\textbf{DP outputs?}}  \\ \hline\hline
Melis et al.\ (2015) \cite{melis_efficient}         & $\cmark$ &          & $\cmark$ & &  &          \\ \hline
Prochlo (2017) \cite{bittau2017prochlo} & \cmark & \cmark & \cmark & \cmark &  & \cmark \\ \hline
P2KMV (2018) \cite{hagen_p2kmv}                     & $\cmark$ & $\cmark$ & $\cmark$ & &  & \cmark         \\ \hline
EDB (2019) \cite{DBLP:journals/popets/AgarwalHKM19} &          & $\cmark$ & $\cmark$ & &  & \cmark        \\ \hline
Honeycrisp (2019) \cite{roth2019honeycrisp} & \cmark &  & \cmark &  & \cmark & \cmark \\ \hline
SANNS (2020) \cite{DBLP:conf/uss/0030CDPRR20}       &          & $\cmark$ & $\cmark$ & $\cmark$ &  &  \\ \hline
Crypt$\epsilon$ (2020) \cite{DBLP:conf/sigmod/ChowdhuryW0MJ20} & \cmark & \cmark & \cmark &  &  & \cmark \\ \hline
Boneh et al.\ (2021) \cite{boneh_lightweight}       & $\cmark$ &          & $\cmark$ & & \cmark & \cmark         \\ \hline
Waldo (2022) \cite{DBLP:conf/sp/DautermanRPS22}     & $\cmark$ &          & $\cmark$ &  & \cmark &           \\ \hline
HERS (2022) \cite{hers}                             &          & $\cmark$ & $\cmark$ & $\cmark$ &   &  \\ \hline
Whisper (2023) \cite{whisper}                       & $\cmark$ &          & $\cmark$ &  & \cmark & \cmark        \\ \hline
\textbf{\Synopsis (this work)}                                 & $\cmark$ & $\cmark$ & $\cmark$ & $\cmark$ & $\cmark$ & $\cmark$ \\ \hline \hline
\end{tabular}
} 
\caption{Works offering cryptographic protection to data that remains secret to a querier (aside from query output).}
\label{tab:relatedworks}
\end{table}
\section{Preliminary Definitions}
\label{sec:preliminaries}

$\Vert \cdot \Vert$ denotes $\ell_2$ distance.  $\Vert \cdot \Vert_1$ denotes $\ell_1$ distance. $\Dist$ denotes cosine distance, defined below in \Cref{def:cosine-distance}.
Given a function $f : \mathcal{R}^n \rightarrow \mathcal{R}^k$, we let $\textsf{Sen}_1(f)$ be the $\ell_1$ sensitivity of $f$, that is, $\textsf{Sen}_1(f) = \max_{X,Y \in \mathcal{N}^n, \Vert X-Y \Vert_1 = 1} \Vert f(X) - f(Y) \Vert_1$. Let $\textsf{Sen}_2(f)$ be analogously defined for $\ell_2$ sensitivity.
If $X$ is a matrix, we say $Y$ is ``neighboring'' to $X$ if $Y$ and $X$ differ in at most one row.
We note that ``count queries,'' in which $f(X)$ returns a count of rows within $X$ that match an arbitrary predicate, have $\textsf{Sen}_1(f) = 1$.
$\text{Lap}(b)$ represents the Laplace distribution with parameter $b$, that is the distribution $\frac{1}{2b} \text{exp}(-\vert x \vert/b)$.
$\mathcal{N}(\mu, \sigma^2)$ represents a normal distribution with mean $\mu$ and variance $\sigma^2$.

\begin{definition}[$(\epsilon, \delta)$-Differential Privacy (DP; adapted from \cite{dwork_2006})]
\label{def:dp}
A randomized algorithm $\CMcal{M}$ with domain $[0,1]^{\NumMsgs \times \EmbLen}$ is $(\epsilon, \delta)$-differentially private if for all $\CMcal{S} \subseteq \textrm{Range}(\CMcal{M})$ and for all neighboring $X, Y \in [0,1]^{\NumMsgs \times \EmbLen}$,
$$ \Pr[\CMcal{M}(x) \in \CMcal{S}] \leq \textrm{exp}(\epsilon) \Pr[\CMcal{M}(y) \in \CMcal{S}] + \delta.$$
\end{definition}

\begin{definition}[Cosine similarity and cosine distance (\cite{DBLP:journals/coling/Xiong16} Def.\ 2.4)]
\label{def:cosine-distance}
The \emph{cosine similarity} between two vectors $a$ and $b$ is $\textsf{CosSim}(a,b) = \frac{1}{\Vert a \Vert \Vert b \Vert} (a \cdot b)$ (where $\cdot$ is the dot product), which is also the cosine of the angle between $a$ and $b$.

We define \emph{cosine distance} as $\Dist(a,b) = 1 - \textsf{CosSim}(a,b)$.
\end{definition}
Cosine similarity is a standard notion of distance between semantic vector embeddings in natural language processing \cite{DBLP:journals/coling/Xiong16}).
See \Cref{app:cosine-facts} for additional details on cosine distance.

\begin{definition}[Laplace Mechanism (\cite{dwork_algorithmic_2013} Def.\ 3.3)]
\label{def:laplace}
Given any function $f : \mathcal{N}^n \rightarrow \mathcal{R}^k$, the \emph{Laplace mechanism} is defined as $f(x) + (Y_1, \ldots, Y_k)$ where $Y_i$ are i.i.d. random variables drawn from $\text{Lap}(\textsf{Sen}(f)/\epsilon)$.  The Laplace mechanism is $(\epsilon,0)$ differentially private (\cite{dwork_algorithmic_2013} Thm.\ 3.6).
\end{definition}

\begin{definition}[Sparse Vector Mechanism (\cite{smith_svt} Def.\ 2, adapted from \cite{DBLP:conf/stoc/DworkNRRV09})]
\label{def:svt}
Let $X$ be a matrix, let $T$ be a threshold, and let $f_1$, $f_2$, $\ldots$ be a stream of queries over the rows of $X$. Then the \emph{sparse vector mechanism} is defined as the following algorithm.  Let $\tilde{T} = T + \text{Lap}(2 \cdot\textsf{Sen}(f)/\epsilon)$. Starting at $i=0$, let $\tilde{a}_i = q_i(X) + \text{Lap}(4 \cdot \textsf{Sen}(f)/\epsilon)$.  If $\tilde{a}_i < \tilde{T}$ then output False and repeat with $i$ incremented by 1. Else, output True and exit.
This algorithm is $(\epsilon,0)$-differentially private (\cite{smith_svt} Thm.\ 1). 
\end{definition}

\begin{definition}[Gaussian Mechanism (\cite{dwork_algorithmic_2013}, adapted from \cite{DBLP:conf/eurocrypt/DworkKMMN06})]
\label{def:gaussian}
Given any function $f : \mathcal{N}^n \rightarrow \mathcal{R}^k$, the \emph{Gaussian mechanism} is defined as $f(x) + (Y_1, \ldots, Y_k)$ where $Y_i$ are i.i.d. random variables drawn from $\mathcal{N}(0, \textsf{Sen}_2(f) \sqrt{2\ln(1.25/\delta)}/\epsilon)$.  The Gaussian mechanism is $(\epsilon,0)$ differentially private (\cite{dwork_algorithmic_2013} Thm.\ A.1).
\end{definition}

\begin{definition}[PrivateProjection Mechanism (\cite{Kenthapadi_2013} Alg.\ 1)]
\label{def:privateprojection}
Let $X$ be an $n \times d$ matrix with entries between 0 and 1, let parameters $\epsilon, \delta$, and dimensions $k$.
The \emph{PrivateProjection mechanism} is the following algorithm.
Let $P$ be a random $d \times k$ projection matrix satisfying Johnson-Lindenstrauss guarantees \cite{Kenthapadi_2013,lindenstrauss1984extensions}.  Output $XP + \Delta$, where $\Delta$ is a matrix where each element is sampled from $\mathcal{N}(0,\sigma^2_\Delta)$, where $\sigma_\Delta$ is a function of $\epsilon$, $\delta$, and $P$ given in \Cref{eqn:noise-matrix-sigma}.
PrivateProjection is $(\epsilon,\delta)$-differentially private (\cite{Kenthapadi_2013} Thm.\ 1).
\end{definition}

\begin{definition}[Malicious-secure Multi-Party Computation (MPC; adapted from \cite{evans2018pragmatic})]
A protocol $\Pi$ securely realizes ideal functionality $\mathcal{F}$ in the presence of malicious adversaries if for every real-world computationally-bounded adversary $\mathcal{A}$ there exists a simulator $\mathcal{S}$ with $\textsf{corrupt}(\mathcal{A}) = \textsf{corrupt}(\mathcal{S})$ such that, for all inputs for honest parties $\{x_i : i \notin \textsf{corrupt}(\mathcal{A})\}$, we have $\textsf{Real}_{\Pi,\mathcal{A}}(\{x_i : i \notin \textsf{corrupt}(\mathcal{A})\}) \approx_c \textsf{Ideal}_{\mathcal{F},\mathcal{S}}(\{x_i : i \notin \textsf{corrupt}(\mathcal{S})\})$.
\end{definition}
Informally, an MPC protocol among $\NParties$ parties ``realizes an ideal functionality'' if it ensures that no party gains ``extra'' information from the real-world protocol beyond what the ideal functionality would provide, under cryptographic assumptions.
\section{Methodology for Privacy Architecture Choices}
\label{sec:methodology}
\label{sec:design}

In this section, we describe our analysis of stakeholder privacy and security concerns in the context of journalistic investigations of WhatsApp messages. We use that analysis to formulate technical goals and identify design priorities.
The final system, \Synopsis, supports versatile journalistic analysis of E2EE chats at the trend level only, while providing strong privacy guarantees for data donors and all members of their chats.

We remark that \Synopsis was designed in the context of an existing tool, WhatsAppWatch, deployed by the data stewards mentioned in \Cref{sec:stakeholder}, for donors to automatically donate a subset of their WhatsApp texts.
That tool is already used to support journalistic investigations.
This project began as that organization was investigating how it should ethically and practically treat donated WhatsApp messages from private chats, and sought to provide the highest level of privacy to potentially-revealing messages that would still enable useful analysis.

\subsection{Stakeholder analysis}
\label{sec:stakeholder}
\label{sec:stakeholder-analysis}

\newcommand{\Journalist}{Journalist\xspace}
\newcommand{\Journalists}{\Journalist{s}\xspace}
\newcommand{\ServerController}{Server Controller\xspace}
\newcommand{\ServerControllers}{\ServerController{s}\xspace}
\newcommand{\DataDonor}{Data Donor\xspace}
\newcommand{\DataDonors}{\DataDonor{s}\xspace}
\newcommand{\OtherUser}{Other User\xspace}
\newcommand{\OtherUsers}{\OtherUser{s}\xspace}
\newcommand{\Public}{Public\xspace}

We begin by identifying the parties involved in \Synopsis: \Journalists, \ServerControllers, \DataDonors, \OtherUsers, and the \Public.

\paragraph{\Journalists (Queriers, Clients).}
\Journalists are the parties that make queries to the \Synopsis database.
Their ultimate goal is to \textbf{obtain useful journalistic insights}.
We identified three key needs for this system that would be necessary for this to be the case:
\begin{enumerate}
\item Timely access to information about events unfolding quickly (e.g. within hours), including ``superspreading'' messaging.
\item Ability to perform retrospective investigations (e.g. years after an event occurred).
\item Results that are reasonably accurate in a two-sided way---minimizing false leads while also ensuring that key insights aren’t missed.
\end{enumerate}
We remark that, for reasons discussed further in \Cref{sec:why-mpc}, it is technically challenging to enable immediate access to information \textit{and} post hoc access to message corpora many weeks or month after initial collection.

In addition to these needs, we also identified some non-issues that allowed us to relax certain design constraints.
In particular, we found that \emph{trend data is sufficient for many kinds of journalistic insights}. Though more granular information is rarely unwelcome in journalism, less granular trends have proven sufficiently insightful while helping balance the ethical responsibility to minimize harm.

\paragraph{\ServerControllers (Data Stewards)}
The designers of WhatsApp Watch already serve as data stewards for a large volume of donated WhatsApp data, and maintain both the architecture that enables message donations and the data itself. These data stewards have ongoing relationships with journalists and data donors on the ground, and are deeply invested in maintaining long-term trust with both communities. 

With the ultimate goal of building a robust system to enable hard-hitting investigative data journalism, data stewards aim to \emph{make clear, credible commitments} about how data is stored and used, and \emph{back their claims up with evidence.}

In order to build a robust system, data stewards must maintain long-term trust with both journalists and data donors. This starts with confidently being able to say, ``We do not, and cannot, misuse your data.'' In particular, the system must prioritize usefulness for investigative journalism and also:

\begin{enumerate}
    \item Minimize access to raw messages, even for the data stewards themselves.
    \item Minimize the risk of leaks, subpoenas, and re-identification, where possible.
    \item Minimize infrastructural and computational costs.
\end{enumerate}

\paragraph{\DataDonors}
Donors voluntarily contribute messages and message  history of a chosen subset of their WhatsApp groups.  They would like to \textbf{avoid negative repercussions of data donation}. This imposes the following constraints on the system:
\begin{enumerate}
\item Donation should be consensual and informed
\item Donation should be easy and smooth
\item Guarantees about data storage and use must be clear, precise, and demonstrably robust. 

\end{enumerate}
The security and privacy concerns of the data donors were the highest priority for us---\textbf{if this system is going to work at all, it must protect its sources}.
As we will discuss in \Cref{sec:approaching-e2ee}, the primary goal of our system is to provide this privacy and security as much as possible while still allowing these sources to make donations with a significant external social benefit.

\paragraph{\OtherUsers}
These are WhatsApp users who share chats with data donors, so their data appears in the database even though they did not consent to the donation itself (similar to how many journalistic sources report on information about people without their consent).  They are not parties in the protocol itself.
We still wish to preserve their privacy as much as possible, with the same protection as the data donors' sent messages.
Thus our goal for this class of users is essentially \textbf{to preserve the privacy of their individual messages} as much as possible within other constraints.

As we will discuss in \Cref{sec:unit-of-privacy-superspreaders}, it may be possible to glean some information about a particular type of user from the output of the system: namely, superspreaders whose messages are high-volume matches for journalist queries.
This visibility is a tradeoff in service of the usefulness of the system, but we emphasize that the protection afforded by our system is a \emph{significant} improvement over the protection of these users' messages in other WhatsApp text-monitoring systems, where only lightly-redacted raw messages appear in databases in the clear.
Research on messaging behavior indicates that most users send significantly fewer messages than these high-volume users \cite{kalogeropoulos2023unraveling,tan2021tracking}. Therefore, we design \Synopsis to enable measurement of these high-volume users without impacting the vast majority of users (and while still providing strong privacy even to the minority who do send many messages).

\paragraph{\Public}
The public is not a party of this system in a cryptographic sense; however, there are significant public interest issues at play.
In some cases, misinformation spread on WhatsApp has directly led to vigilante killings and other incidents that have negatively impacted members of the public (e.g. \cite{shekhar_guns_2024,lankesh1,lankesh2,ram_scam_2024}). 
Simultaneously, we believe that the public would be negatively affected by a system that employed surveillance and non-consensual data collection to perform investigative analysis.
We believe that the public's best interests are served by \emph{ethical}, accurate, and timely journalistic investigations of propaganda and disinformation.
(Like the \OtherUsers, the \Public is also not party to the protocol itself.)

\paragraph{\textbf{Remark}}
This system balances competing design principles in our use case for our current stakeholders.  However, we strongly warn that it cannot blindly be extrapolated to other settings with different stakeholders.
In particular, in our case the underlying messaging provider WhatsApp is not affiliated with the data stewards.
The analysis would be very different when that service provider plays a role in a \Synopsis-like system, in part because service providers have very different incentives than those analyzed here (including but not limited to monetization of the service).

\subsection{Iterative design process summary}

Equipped with this stakeholder analysis, we proceeded to use the following method to design our system:
\begin{enumerate}
\item Perform the stakeholder analysis above to identify the parties and their constraints, needs, and wants
\item Choose priorities for the starting point for our design.  Our two highest priorities were (1) the security and privacy of the data donations, and (2) the ability to obtain useful journalistic insights from the system.  We make two remarks on why these were both necessary conditions for our system:
    \begin{itemize}
    \item   The privacy and security of the the data donations and donors had the highest priority, since (1) they are the main party who the system is built to enable, (2) they are ``paying'' the most into the system with their personal data and assuming the most risk.  \textbf{If this system fails to protect data donor sources, the whole system loses its main appeal to donors, the loss of donors removes its appeal to other parties.}
    \item Enabling journalistic insights was also of the highest priority. As other, less-private data donation mechanisms are introduced, \textbf{journalists will simply use less private alternatives if \Synopsis is not sufficiently useful}. We believe that journalistic insights are positive, in and of themselves, since they significantly benefit the public. The existence of less private alternatives means that this usefulness requirement must be balanced against the privacy requirement, and should not be subordinate to it.
    \end{itemize}
\item With these priorities identified, we created an initial design and began to look for important design levers and choices.
\item Then, in an iterative process, we considered the degree to which the design is meeting the needs of each stakeholder. If a design decision yielded significant negative consequences for that stakeholder, we re-examined our framework to see if alternative designs might yield more amenable tradeoffs.  We also began iterating on implementations of our design during this process.
\item Once we finalized key design decisions, we chose parameters for our system; continued to seek minor improvements; and continued to implement, benchmark, and test the \Synopsis codebase. 
\end{enumerate}

\subsection{Design decisions}
\label{sec:design-decisions}

In this subsection we describe the design decisions we made during the process described in the last subsection.  These decisions, which touched on some surprisingly subtle aspects of cryptographic and privacy-preserving data analysis, are usually not given a full treatment in academic literature that focuses on narrower contributions.

\subsubsection{Almost, but not quite, full E2EE guarantees after donation.}
\label{sec:approaching-e2ee}
End-to-end encryption (E2EE) 
provides a host of security guarantees to the messages sent by the ``end'' users, most relevantly providing a confidentiality guarantee that ensures that no parties but the ends---not even the service provider---can distinguish between sending same-length message contents
\cite{DBLP:conf/sp/UngerDBFPG015,DBLP:conf/acns/VatandasGIK20}.
In the implementation that is the best for users of the messaging system, E2EE guarantees would persist even after an end voluntarily donates messages.

Ignoring cryptographic deniability for a moment (we return to it in \Cref{sec:deniability}), this level of protection cannot be given to query outputs of our system under reasonable modeling assumptions. To see why, consider an adversary doing a version of a chosen plaintext attack (CPA) who can also choose a query to our system to detect a trend.  
Imagine a CPA game in which the adversary may instruct a corrupted user to send either 1000 copies of message $m_1$, or 1000 copies of message $m_2$ of equal length, and will then attempt to determine which the sender sent.
If the adversary can also query \Synopsis for the $m_1$ and $m_2$ messages, it should see a spike in sends of $m_1$ or $m_2$, which it can use to trivially win the CPA game.

The attack also remains winnable with a smaller advantage when a single copy of $m_1$ versus $m_2$ is sent instead of 1000.  In fact, this attack remains winnable from a cryptographic perspective even if \Synopsis, in the manner of differential privacy (DP), bounds the advantage the adversary gains in knowing which message was sent.
The advantage is reduced to a constant factor that depends upon the DP parameter $\epsilon$, but is still significantly larger than the $2^{-\lambda}$ advantage needed to declare this system cryptographically secure.

Still, this does not mean that the effective security in these three scenarios is the same. 
Any system that allows trend analysis over the message database necessarily reduces the level of cryptographic protection as compared to the original E2EE system. However, there is still very good reason to restrict the leakage of this system beyond that of a clearly defined multi-party-computation ideal functionality---to design the ideal functionality in such a way that it verifiably does not leak too much about any individual message, and to verify that the ideal functionality is being adhered to by a malicious security guarantee.
Essentially, in \Synopsis we seek to create a system that is \emph{as close to E2EE-guarantees as possible} while meeting our other design goals.

We emphasize that our system provides significantly more protection \emph{from malicious journalists/researchers and data stewards} than do other systems that analyze donated WhatsApp texts \cite{garimella2024whatsappexplorerdatadonation,whatsapp_monitor}.

\subsubsection{Consensual donation by donors is essential.}
A key goal of both the data donors and the data stewards is that donation of the contributed text messages is consensual and informed.  This also aligns well with the principles of end-to-end encryption, which (as discuss further in this section) typically trust all ``ends'' of the communication but do not enable service providers to learn information about content of messages.

Although the current design includes messages sent by chat participants who did not consent to data donation, it is limited to messages \emph{received} (or sent) by chat participants who consented to the donation.  From an E2EE standpoint, this is among the many actions that an end of the communication can do; from a journalistic standpoint, this is similar to how a source might report on the messages sent to them by some other nonconsenting party; and, from a research perspective, limiting the set of donated messages to those \emph{sent} consensually by a donor would be prohibitively restrictive for drawing decisions from data.  We see this as the best way to balance these needs---especially given the high level of privacy we ultimately offer compared to other methods of text analysis.

\subsubsection{Our query interface}
\label{sec:design-dp}

Thanks to clear communication from WhatsApp Watch designers with working relationships with data donors, we were aware of clear steps to gain informed consent from donor WhatsApp users (see \Cref{fig:waw_dialogue}).
Even though many of them were, in theory, willing to share their message history outright, storing a raw or lightly-redacted message database seemed inconsistent with our previously-mentioned goal of providing \emph{as close to E2EE guarantees as possible} to the donated messages.

After much discussion on what would be useful for journalists, it became apparent that raw text information was not needed---a spike in the popularity of specific topics would be a sufficiently useful signal for a variety of investigations.
This led us to a query mechanism that would surface numeric counts or trends in messages, rather than the messages themselves.
This new focus also gave us a natural candidate for privacy to apply to the ideal functionality: differential privacy.

Although many design decisions remained that we will discuss here, this crystallized some useful properties:
\begin{enumerate}
\item Queriers must already have some idea of what they are looking for before making queries; therefore, the system does not easily facilitate mass surveillance for many high-fidelity queries simultaneously.
\item Although we had decided on a system where cryptography would ensure adherence to the ideal functionality, the use of DP would still ensure formal guarantees directly \emph{via the query interface itself.}
\item Even if all other protections failed, ``local DP'' protections would remain on any data stored in that model (see \Cref{sec:local-and-central}).
\end{enumerate}

\subsubsection{Uniting local DP, central DP, and multi-party computation.}
\label{sec:local-and-central}

On deciding to run a differentially private query mechanism, we still needed to make several choices to arrive at an algorithm that could meaningfully balance the concerns of enabling exploratory research, conducting relatively accurate targeted analysis, and using DP to guarantee strong E2EE-like guarantees.

One such choice is between the central and local models of differential privacy. The \emph{central model} of DP represents a trusted curator that receives the inputs of all users in the clear and produces one combined output that is noised centrally for privacy. In contrast, the local model represents an untrusted curator who receives locally noised inputs from each user and simply combines those to produce an output. From a DP perspective, the users could publish their locally private inputs on a bulletin board and each user would then locally combine the published values to produce the output. In general, an algorithm satisfying central DP would have a much lower loss in accuracy compared to equally private local-DP algorithm for the same problem. Therefore, the choice between the two settings usually represents the choice adversarial model of the implementation.

Interestingly, that is not the case in our setting. By running the DP algorithm under MPC, we shift the responsibility of enforcing the adversarial model to the MPC algorithm. The choice of model instead represents a tradeoff between query flexibility and the accuracy of query responses. 

\begin{figure}[t!]
    \centering
    \resizebox{1\linewidth}{!}{
        \begin{tikzpicture}
        	\begin{axis}[
        		xlabel=$\EpsilonFineCount$ (FC) and $\EpsilonCoarse$ (CC),
        		ylabel=mean absolute error ($\log_{10}$),
                            scale only axis,
                    max space between ticks=40,
                    minor x tick num=1,
                    minor y tick num=2,
                    xmin = -0.2,
                    xmax = 4.2,
                    ymode=log,
                    log basis y={10},
                    major tick length=0.15cm,
                    minor tick length=0.075cm,
                    tick style={semithick,color=black}, 
                    legend style={at={(0.65,0.88)},anchor=west},
                    height = 7 cm,
                    width = 9 cm
                    ]

                \addplot+ [ color=\ColorGraphCC,mark=square*,
                    error bars/.cd, y dir=both,y explicit,
                ]
                coordinates {

                (0.2, 65)   +- (26.96005275, 26.96005275)
        	(0.4, 65.38)   +- (36.07399802, 36.07399802)
        	(0.6, 54.85)   +- (23.82249171, 23.82249171)
        	(0.8, 44.96)   +- (34.34530536, 34.34530536)
        	(1.0, 38.61)   +- (38.23160415, 38.23160415)
                (2.0, 14) +- (27.68092002, 27.68092002)
                (3.0, 3) +- (17.00196067, 17.00196067)
                (4.0, 2.192) +- (18.94700211, 18.94700211)
            };
                \addlegendentry{CC, $k = 500$};
                
                \addplot+ [ color=\ColorGraphFC,mark=square*,
                    error bars/.cd, y dir=both,y explicit,
                ]
                    coordinates {
                    (0.2, 4.387673)   +- (2.74869733860747, 2.74869733860747)
        		(0.4, 3.4723653)   +- (3.633432266, 3.633432266)
        		(0.6, 1.344897452)   +- (2.021824168, 2.021824168)
        		(0.8, 0.7863015)   +- (0.5499064873, 0.5499064873)
        		(1.0, 0.6120557259)   +- (0.4927125966, 0.4927125966)
                    (2.0, 0.3204118259) +- (0.2822517416, 0.2822517416)
                    (3.0, 0.45512223) +- (0.496784492, 0.496784492)
                    (4.0, 0.22331239) +- (0.3003045461, 0.3003045461)
        	};
                \addlegendentry{FC, $k = 500$};

        	\end{axis}
        \end{tikzpicture}
        } 
    \caption{Mean absolute error and standard deviation for $\EpsilonFineCount$ (FC) and $\EpsilonCoarse$ (CC) values.} 
    \label{fig:local-central-accuracy}
\end{figure}
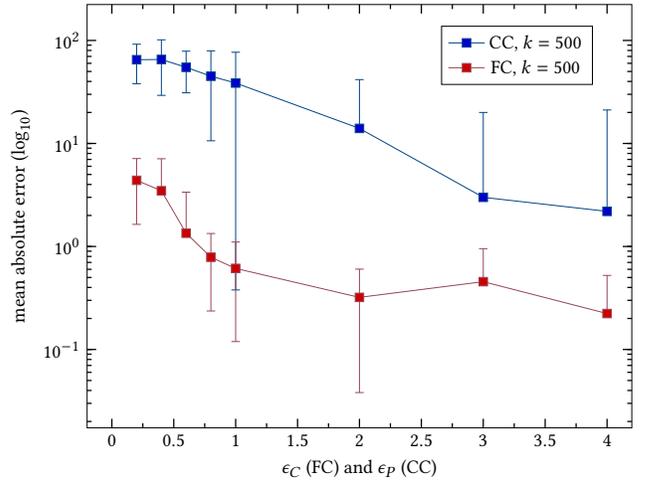

To understand this, consider the local model algorithm where every individual text message embedding is noised to achieve local DP. Since differential privacy is closed under post-processing, this noisy corpus can now be re-used any number of times to check different properties. This would enable exploratory analysis; however, it the error is too high to use such an algorithm in our setting (see \Cref{fig:local-central-accuracy} for an accuracy comparison between \Synopsis queries in the central and local model). In general, the common wisdom is to choose to run the central-DP algorithm in such a case because its privacy-accuracy tradeoff is much better. But this would necessitate giving up on the possibility of exploratory analysis, as journalists would then be restricted to a limited number of queries.

Ultimately, we settled on a hybrid approach. As shown in \Cref{fig:architecture} and described in \Cref{sec:architecture}, we spend some privacy budget on creating a local DP model---creating a ``coarse-grained'' database that can be queried many times but is less accurate.  We then use the rest of the privacy budget on queries to a central DP store, a ``fine-grained'' database that can only be queried a finite number of times per epoch (see \Cref{sec:epoch}). The key insight here relates to the nature of analyses being conducted by journalist queriers, who only require reasonably accurate counts for targeted queries. During exploratory analysis, binary signals---Boolean responses indicating that keyword mentions have exceeded a certain threshold---are sufficient. Even though counts outputs using the coarse database are fairly inaccurate, we show that these counts are still a fairly reliable way to monitor longitudinal topic trends. 

When run under MPC, this ``interface'' forms an MPC ideal functionality. A simplified version of the interface is shown in \Cref{alg:ideal-func}; the full version including MPC simulation details is in \Cref{alg:full-ideal-func}. The interface also allows us to bake in other notions of data privacy alongside DP if desired (e.g. setting a minimum threshold of results necessary for a query to return, similar to cell suppression, which helps ensure this query mechanism is only useful for identifying trends rather than rarely-sent messages), and allows us to restrict queries by rate limiting and access controls.

We remark that our system design deliberately seeks to surmount many of the practical obstacles faced by DP systems \cite{DBLP:journals/corr/abs-2408-07614} through careful query design, use of MPC, clear accounting of privacy budgeting, the ability to explore (less accurate) data without burning budget, and careful consideration of the definition of ``neighboring'' datasets (\Cref{sec:unit-of-privacy}).

\subsubsection{Differential privacy database modeling considerations.}

\paragraph{Definition of ``neighboring'' datasets: those that differ by one donated message}
\label{sec:unit-of-privacy}
\label{sec:unit-of-privacy-superspreaders} 
All differentially private systems must choose how to define the ``neighboring'' databases over which the privacy guarantee holds up to a bound with parameter $\epsilon$.

In our system we consider databases ``neighboring'' if they differ by one text message, however, we could also have designed a system that considers databases ``neighboring'' if they differ in (e.g.) a single word, or a user's entire set of donated messages. 

We choose to provide $\epsilon$ protection to messages rather than users.
This is because some groups on WhatsApp have a small number of users contributing a large number of messages---and in fact these ``superspreaders'' are often the exact parties of specific journalistic interest \cite{tan2021tracking,doi:10.1177/2050157920958442,kalogeropoulos2023unraveling,grinberg2019fake}.  

Unless we choose to deliberately track identity information, we would need to ``over-protect'' users who sent any fewer messages than that large number. This would add a significantly higher level of noise and essentially remove all utility of the system.
We also note that, even with message-level rather than user-level protections, DP does not completely stop working on groups of messages---it simply provides protection to groups of messages at a reduced-privacy $\epsilon$ through standard group privacy rules \cite{dwork_algorithmic_2013,dwork_calibrating_nodate}.

\paragraph{The scope of a query}
\label{sec:epoch}

Since we are using time-series data which arrives over time, and at least some portion of our design involves central DP-style queries that ``spend'' a privacy budget that will eventually be used up, we must carefully consider the differentially private algorithm we use. 

One option would be to store the messages with epoch labels in one large corpus, with each query specifying both a keyword and a time-interval. Even though this algorithm would provide extremely accurate counts for any length of time interval-- it would significantly complicate the process of tracking the evolving privacy budget. Instead, in order to maintain both simplicity and flexibility, we divide our data up by epochs, and restrict journalists to request a count for a single query on a single epoch (day) at a time.

Given this algorithm, one can treat queries that cover multiple epochs as separate queries to those epochs, and then combine the outputs. 
A key insight for analysis of this algorithm is to remember that our notion of privacy is over datasets that differ in a single text message. This means that asking this algorithm for multiple outputs only expends higher privacy budget when multiple overlapping queries are made over the same epoch.

\subsubsection{Deniability and integrity against corrupt donors.}
\label{sec:deniability}
The security guarantess we can make against a corrupt data donor are subtle.

The only way in which a donor may attempt to break integrity is to alter its input---in other words every possible donor message is consistent with \emph{some} valid input to the cryptographic protocol.
It is therefore tempting to conclude that we have security against a malicious donor.
However,  we do rely on the donor client to submit messages with a specific format (e.g. shares of a norm-1 $k$-vector) and that came from a specific source (groups on WhatsApp).

Due to the by-design deniability property of E2EE WhatsApp messages \cite{fbsecretconversations}, 
the receiver of a WhatsApp message cannot prove that the donated message was actually received by the sender (versus the receiver simulating it independently).
This property cannot be changed without altering WhatsApp's end-to-end encryption protocol.  Furthermore, deniability is often a core goal of E2EE systems \cite{DBLP:conf/sp/UngerDBFPG015,double_ratchet,fbsecretconversations} that we want to keep intact for \Synopsis donors.

However, we do take steps to prevent or detect malicious donor behavior.  The donation application is implemented as a custom WhatsApp client; the messages the client forwards to the \Synopsis servers were received by the custom client from WhatsApp servers.  In order to send false messages to the \Synopsis servers, a corrupt donor would need to essentially write a second custom WhatsApp client to spoof messages from a legitimate donor client.  Furthermore, if this was detected, the \Synopsis servers could refuse to accept future messages from the corrupt client. 

We therefore consider \Synopsis to offer malicious security against UI-bound adversaries \cite{DBLP:conf/chi/FreedPMLRD18} but only semi-honest security (i.e. no integrity) against fully active non-UI-bound adversaries.

\subsubsection{Additional points}

See \Cref{app:additional-design} for additional design points, including our choice of MPC compared with functional or homomorphic encryption (\Cref{sec:why-mpc}), malicious security (\Cref{sec:malicious-security}), and the non-privacy of journalists' queries to \Synopsis (\Cref{sec:query-privacy}).

\section{Technical Architecture}
\label{sec:architecture-overview}
\label{sec:architecture}

Our primary technical goal is to return privacy-preserving trend lines over an epoch range that reflects the \emph{number of matches per epoch that are within a certain distance of a query message in the chosen embedding's semantic space}, as demonstrated by the trend line plotted in \Cref{fig:trendlines}.

The most natural way to do this would be to perform count queries for matches within each epoch.
One of our query mechanisms (fine-grained count queries, \textbf{FC}) does exactly this, using the Laplace mechanism under MPC to provide DP counts at each epoch. Each time this query is made for a particular epoch, this mechanism spends $\EpsilonFineCount$ budget for that data point.
However, we expect that most queries  would not necessarily reveal interesting trends, so using this mechanism to perform those queries would``waste'' privacy budget. 

To avoid this outcome, we provide two different ways to make ``business as usual'' queries cheaper: 
\begin{enumerate}
\item \textbf{Threshold rather than count queries.} Rather than receiving exact counts of matches via the Laplace mechanism (burning $\EpsilonFineCount$), the journalist only learns whether the count is above or below a threshold via the Sparse Vector mechanism \cite{DBLP:conf/stoc/DworkNRRV09,smith_svt}.  This query type burns no privacy budget as long as the count remains below a threshold, and then burns $\EpsilonFineThreshold$ once crossed, creating longstanding queries that run ``for free'' until a high volume of texts are sent on a particular topic, at which point the journalist could turn to more detailed count queries, acting as an "alert" system for them.
\item \textbf{Querying noisier synthetic data generated in the local DP model (``coarse-grained'') rather than exact data in the central DP model (``fine-grained'').}
This requires a one-time privacy cost of $\EpsilonCoarse$ to build a perturbed dataset using the PrivateProjection mechanism (a variant of the Gaussian mechanism; see \Cref{def:privateprojection} and  \cite{Kenthapadi_2013}). Once that data set is built, it can be queried an arbitrary number of times without expending any privacy budget.
To enable these queries we store a second ``perturbed'' version of the dataset.
\end{enumerate}

These two relaxations can be combined;
in total, we create four query types that rely on three DP mechanisms.
This architecture is summarized in \Cref{fig:queries-to-mechanism} and is described in more detail in this section.
Note that query abbreviations use two letters: the first (\textbf{F/C}) indicates granularity as fine (central DP) or coarse (local DP); the second (\textbf{C/T}) indicates the output as count or threshold.

To serve both coarse- and fine-grained queries, we maintain two copies of the semantic vector database. Once the fine-grained query budget for an epoch runs out, we delete the fine-grained version of that dataset, leaving only the coarse-grained (local DP) dataset.

 \begin{figure}[t!]
     \centering
     \resizebox{\linewidth}{!}{
     \begin{tikzpicture}

\def\xServers{2.5}
\def\wServer{0.6}
\def\hServerSeg{0.3}
\def\dServers{1.7}
\def\yServer{0}
\def\yServerCenter{\yServer}

\def\wCloudLimited{7}
\def\hCloudLimited{5}
\def\xCloudLimited{7}
\def\yCloudLimited{0}

\def\dPhones{1.1}
\def\xPhones{-1}
\def\yPhonesCenter{-0.5}
\def\ysPhones{{\yPhonesCenter-2*\dPhones,\yPhonesCenter-\dPhones, \yPhonesCenter,\yPhonesCenter+\dPhones,\yPhonesCenter+2*\dPhones}}

\def\xaArrowPS{\xPhones+0.5}
\def\xbArrowPS{\xServers-1.3*\wServer}
\def\dyArrowP{\dPhones/1.1/2}
\def\yasArrowPS{{\yPhonesCenter-2*\dPhones+\dyArrowP,\yPhonesCenter-\dPhones+\dyArrowP, \yPhonesCenter+\dyArrowP, \yPhonesCenter+\dPhones+\dyArrowP, \yPhonesCenter+2*\dPhones+\dyArrowP}}
\def\ybsArrowPS{\yServer}

\def\xdsCircles{{0.8, 1.1, 1.5}}
\def\xdsCirclesCenter{{0.7, 0.92, 1.28}}
\def\ydsCircles{{1.34, 1.3, 1.5}}
\def\rsCircles{{0.05, 0.1, 0.2}}

\def\xJourno{15}
\def\yJourno{0}
\def\rJourno{1}

\def\xaLine{\xCloudLimited-3}
\def\xbLine{\xJourno-1}
\def\yLine{\yServerCenter}

\node[cloud, cloud puffs=10, draw, minimum width=\wCloudLimited*0.75 cm, minimum height=\hCloudLimited*0.7 cm, color=black] at (\xCloudLimited, \yCloudLimited) {};

\node[database, database radius=\wServer cm, database segment height=\hServerSeg cm, color=black] at (\xServers,\yServer) {};

\foreach \i in {0, 1, 2} {
    \draw[color=black] (\xServers + \xdsCirclesCenter[\i], \yServerCenter) circle (\rsCircles[\i]); 
}

\def\dataPointsRadius{1.5}
\def\dataPointsBotN{11}
\def\dataPointsBotX{{   0.32, 0.25, 0.2, 0.3, 0.65, 0.4, 0.4, 0.5, 0.55, 0.57, 0.65, 0.7}}
\def\dataPointsBotY{{     -0.1, -0.04, 0.2, 0.44, -0.15,  0.3, 0.6, 0.4, -0.28,  0.1,  0.3,  0.2}}
\def\dataPointsBotDists{{ 0.2, 0.2,  0.3, 0.2, 0.4,  0.3, 0.4, 0.4, 0.4,  0.3,  0.3,  0.3}}
\def\dataPointsBotAngles{{200, 290,  300, 10,  210,  200, 10,  250, 150,  40, 80,    330}}
\def\queryPointRadius{3}
\def\mechHighlightRadius{3.2}
\def\queryPointBotX{0.7}
\def\queryPointBotY{0.65}
\def\queryRadius{20}

\foreach \i in {0, 1, ..., \dataPointsBotN} {
    \pgfmathparse{\dataPointsBotX[\i]*\wCloudLimited + \xCloudLimited - \wCloudLimited/2} \let\x\pgfmathresult
    \pgfmathparse{\dataPointsBotY[\i]*\hCloudLimited/2 - \hCloudLimited/10} \let\y\pgfmathresult
    \pgfmathparse{\dataPointsBotDists[\i]} \let\pertDist\pgfmathresult
    \coordinate (Orig) at (\x cm, \y cm);
    \draw[fill = black, draw = black, line width=0pt] (Orig) circle (\dataPointsRadius pt);
}

\pgfmathparse{\queryPointBotX*\wCloudLimited + \xCloudLimited - \wCloudLimited/2 - 0.2} \let\qbx\pgfmathresult \pgfmathparse{\queryPointBotY*\hCloudLimited/2 - \hCloudLimited/2 + 1} \let\qby\pgfmathresult
\filldraw[\ColorQuery] (\qbx, \qby) circle (\queryPointRadius pt);
\node[anchor=center] (QueryPoint) at (\qbx, \qby) {};
\draw[dashed, \ColorQuery] (\qbx, \qby) circle (\queryRadius pt);

\def\splitFineFig{1 cm}

\node[anchor=center, xshift=2.5cm, yshift=\splitFineFig, text width=1.8cm, align=center] (TextCount) at (\qbx, \qby) {Count of matching points};
\node[anchor=center, xshift=2.5cm, yshift=-\splitFineFig, text width=1.8cm, yshift=-0.1cm, align=center] (TextThreshold) at (\qbx, \qby) {Threshold};
\node[align=center, yshift=-1.5cm, text width=2.7cm] at (\xServers, \yServer) (TextFineDB) {Fine-grained database (donated vector embeddings)};
\node[align=center, yshift=-2.3cm, text width=5cm] at (\xCloudLimited, \yCloudLimited) (TextCloud) {Exact text embeddings for selected time window};

\node[MechanismBox, text width=2.5cm, right of=TextCount, xshift=2.5cm, color=\ColorLapNoise] (MechanismBoxLaplace) {Laplace Mechanism ($\EpsilonFineCount$/query; \Cref{def:laplace})};
\node[MechanismBox, text width=2.5cm, below of=MechanismBoxLaplace, yshift=-0.83cm, color=\ColorLapNoise] (MechanismBoxSparse) {Sparse Vector Mechanism ($\EpsilonFineThreshold$/query; \Cref{def:svt})};
\path (MechanismBoxSparse.north west) -- (MechanismBoxSparse.south west) coordinate[pos=0.33] (MBSTopLeft);
\path (MechanismBoxSparse.north west) -- (MechanismBoxSparse.south west) coordinate[pos=0.67] (MBSBotLeft);

\draw[->] ({\qbx + 0.65},{\qby + 0.3}) -- (TextCount.west);
\draw[->] ({\qbx + 0.65},{\qby - 0.3}) -- (TextThreshold.west);

\node[QueryBox, right of=MechanismBoxLaplace, xshift=2.2cm, text width=2.4cm, align=center, color=\ColorQuery] (QueryFC) {Fine-grained Count Query (FC; \Cref{alg:query-fine-count})};
\node[QueryBox, align=center, right of=MechanismBoxSparse, xshift=2.2cm, text width=2.4cm, color=\ColorQuery] (QueryFT) {Fine-grained Threshold Query (FT; \Cref{alg:query-fine-threshold})};

\draw[->] (TextCount) -- (MechanismBoxLaplace);
\node[right of=TextCount, yshift=0.1cm, xshift=0.5cm] (ArrowSplitPoint) {};
\draw[->] (ArrowSplitPoint) |- (MBSTopLeft);
\draw[->] (TextThreshold) -- (MBSBotLeft);

\draw[->] (MechanismBoxLaplace) -- (QueryFC);
\draw[->] (MechanismBoxSparse) -- (QueryFT);

\end{tikzpicture}}
     \caption{Fine-grained query types.  Either report a noised count of eligible matches (\textbf{FC}) or suppress below-threshold query results \textbf{(FT)}. Both queries expend privacy budget.}
     \label{fig:enter-label-fg}
 \end{figure}
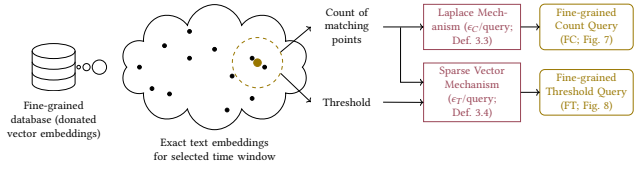
 
 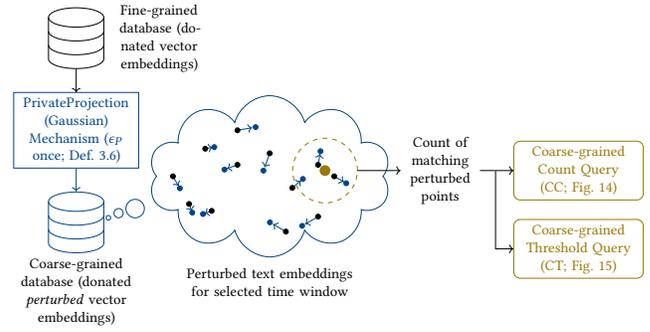
\begin{figure}[t!]
     \centering
     \resizebox{\linewidth}{!}{
     \begin{tikzpicture}

\def\xServers{2.8}
\def\wServer{0.6}
\def\hServerSeg{0.3}
\def\dServers{1.7}
\def\yServer{-1}
\def\yServerCenter{\yServer}

\def\wCloudLimited{7}
\def\hCloudLimited{5}
\def\xCloudLimited{7}
\def\yCloudLimited{0}

\def\dPhones{1.1}
\def\xPhones{-1}
\def\yPhonesCenter{-0.5}
\def\ysPhones{{\yPhonesCenter-2*\dPhones,\yPhonesCenter-\dPhones, \yPhonesCenter,\yPhonesCenter+\dPhones,\yPhonesCenter+2*\dPhones}}

\def\xaArrowPS{\xPhones+0.5}
\def\xbArrowPS{\xServers-1.3*\wServer}
\def\dyArrowP{\dPhones/1.1/2}
\def\yasArrowPS{{\yPhonesCenter-2*\dPhones+\dyArrowP,\yPhonesCenter-\dPhones+\dyArrowP, \yPhonesCenter+\dyArrowP, \yPhonesCenter+\dPhones+\dyArrowP, \yPhonesCenter+2*\dPhones+\dyArrowP}}
\def\ybsArrowPS{\yServer}

\def\xdsCircles{{0.8, 1.1, 1.5}}
\def\xdsCirclesCenter{{0.7, 0.92, 1.28}}
\def\ydsCircles{{0.1, 0.2, 0.3}}
\def\rsCircles{{0.05, 0.1, 0.2}}

\def\xJourno{15}
\def\yJourno{0}
\def\rJourno{1}

\def\xaLine{\xCloudLimited-3}
\def\xbLine{\xJourno-1}
\def\yLine{\yServerCenter}

\node[cloud, cloud puffs=10, draw, minimum width=\wCloudLimited*0.75 cm, minimum height=\hCloudLimited*0.7 cm, color=\ColorPerturbed] at (\xCloudLimited, \yCloudLimited) {};

\node[database, database radius=\wServer cm, database segment height=\hServerSeg cm, color=\ColorPerturbed] (CoarseDB) at (\xServers,\yServer) {};

\foreach \i in {0, 1, 2} {
    \draw[color=\ColorPerturbed] (\xServers + \xdsCirclesCenter[\i], \yServerCenter + \ydsCircles[\i]) circle (\rsCircles[\i]); 
}

\def\dataPointsRadius{1.5}

\def\dataPointsBotN{11}
\def\dataPointsBotX{{   0.32, 0.25, 0.2, 0.3, 0.65, 0.4, 0.4, 0.5, 0.55, 0.57, 0.65, 0.7}}
\def\dataPointsBotY{{     -0.1, -0.04, 0.2, 0.44, -0.15,  0.3, 0.6, 0.4, -0.28,  0.1,  0.3,  0.2}}
\def\dataPointsBotDists{{ 0.2, 0.2,  0.3, 0.2, 0.4,  0.3, 0.4, 0.4, 0.4,  0.3,  0.3,  0.3}}
\def\dataPointsBotAngles{{200, 290,  300, 10,  210,  200, 10,  250, 150,  40, 80,    330}}
\def\queryPointRadius{3}
\def\mechHighlightRadius{3.2}
\def\queryPointBotX{0.7}
\def\queryPointBotY{0.65}
\def\queryRadius{20}

\foreach \i in {0, 1, ..., \dataPointsBotN} {
    \pgfmathparse{\dataPointsBotX[\i]*\wCloudLimited + \xCloudLimited - \wCloudLimited/2} \let\x\pgfmathresult
    \pgfmathparse{\dataPointsBotY[\i]*\hCloudLimited/2 - \hCloudLimited/10} \let\y\pgfmathresult
    \pgfmathparse{\dataPointsBotDists[\i]} \let\pertDist\pgfmathresult
    \coordinate (Orig) at (\x cm, \y cm);
    \coordinate (Pert) at ($(Orig) +(\dataPointsBotAngles[\i]:\pertDist cm)$);
    \draw[->, \ColorPerturb] (Orig) -- ($(Orig)!0.8!(Pert)$);
    \draw[fill = black, draw = black, line width=0pt] (Orig) circle (\dataPointsRadius pt);
    \draw[fill = \ColorPerturb, draw = \ColorPerturb, line width=0pt] (Pert) circle (\dataPointsRadius pt);
}

\pgfmathparse{\queryPointBotX*\wCloudLimited + \xCloudLimited - \wCloudLimited/2 - 0.2} \let\qbx\pgfmathresult \pgfmathparse{\queryPointBotY*\hCloudLimited/2 - \hCloudLimited/2 + 1} \let\qby\pgfmathresult
\filldraw[\ColorQuery] (\qbx, \qby) circle (\queryPointRadius pt);
\node[anchor=center] (QueryPoint) at (\qbx, \qby) {};
\draw[dashed, \ColorQuery] (\qbx, \qby) circle (\queryRadius pt);

\node[anchor=center, xshift=2.5cm, text width=1.5cm, align=center] (TextCountPerturbed) at (\qbx, \qby) {Count of matching perturbed points};
\node[align=center, yshift=-1.5cm, text width=2.5cm] at (\xServers, \yServer) (TextCoarseDB) {Coarse-grained database (donated \emph{perturbed} vector embeddings)};
\node[align=center, yshift=-2.3cm, text width=5cm] at (\xCloudLimited, \yCloudLimited) (TextPerturbedCloud) {Perturbed text embeddings for selected time window};

\draw[->] ({\qbx + 0.7},\qby) -- (TextCountPerturbed);

\node[QueryBox, right of=TextCountPerturbed, xshift=2cm, text width=2.6cm, align=center, color=\ColorQuery] (QueryCC) {Coarse-grained Count Query (CC; \Cref{alg:query-coarse-count})};
\node[QueryBox, below of=QueryCC, yshift=-0.7cm, text width=2.6cm, align=center, color=\ColorQuery] (QueryCT) {Coarse-grained Threshold Query (CT; \Cref{alg:query-coarse-threshold})};

\draw[->] (TextCountPerturbed) -- (QueryCC);
\node[right of=TextCountPerturbed, yshift=0.1cm, xshift=0.2cm] (ArrowSplitPoint) {};
\draw[->] (ArrowSplitPoint) |- (QueryCT);

\node[MechanismBox,above of=CoarseDB, color=\ColorGaussPerturb, yshift=1cm, minimum width=2.5cm, align=center, text width=2.5cm] (MechanismBoxGaussian) {PrivateProjection (Gaussian) Mechanism ($\EpsilonCoarseTotal$ once; \Cref{def:privateprojection})};
\node[database, database radius=\wServer cm, database segment height=\hServerSeg cm, color=black, above of=MechanismBoxGaussian, yshift=1cm] (FineDB) {};
\node[align=center, right of=FineDB, xshift=0.8cm, yshift=4cm, text width=2.5cm] at (\xServers, \yServer) (TextCoarseDB) {Fine-grained database (donated vector embeddings)};

\draw[->] (FineDB) -- (MechanismBoxGaussian);

\draw[->, color=\ColorGaussPerturb] (MechanismBoxGaussian) -- (CoarseDB);

\end{tikzpicture}}
     \caption{Coarse-grained query types.  Either report the count \textbf{(CC)} or a Boolean true/false statement of whether or not the count is above a threshold \textbf{(CT)} on the perturbed data.} 
     \label{fig:enter-label-cg}
     \vspace{-0.5em}
 \end{figure}

\begin{figure}[h]
\fbox{%
\begin{minipage}{0.97\linewidth}
\footnotesize
\textbf{Simplified ideal functionality for \Synopsis (Omits some MPC simulation details; full version in \Cref{alg:full-ideal-func}.)} \\
\textbf{Parties}: \emph{Donors} submit input messages, \emph{Servers} controlled by the data stewards store shares of the fine-grain and coarse-grain dataset of messages, \emph{Journalists} make queries and receive outputs of queries. \\
\textbf{Setup}:
\begin{itemize}[left=0.1cm]
\item Set maximum per-epoch fine-grain budget $\EpsilonFineTotal$.
\item Set coarse-grain budget $\EpsilonCoarseTotal$, generate projection matrix $P$ from $\EpsilonCoarseTotal$ (see \Cref{alg:kkmm}).
\item When reaching a new epoch $\Epoch$, initialize empty coarse database $\DBCoarseEpoch$, and initialize empty fine database $\DBFineEpoch$ with budget $\EpsilonFEpoch = \EpsilonFineTotal$.
\end{itemize}
\textbf{Upon receiving a new donated message $m$ during epoch $\Epoch$ from a donor}:
\begin{itemize}[left=0.1cm]
\item Get message embedding $x$ from $m$.  Store $x$ in the fine-grain database $\DBFineEpoch$.
\item Compute perturbed embedding $\tilde{x} = x + r$ (for multi-variate Gaussian $r$) as described in \Cref{alg:storage}. Store $\tilde{x}$ in the coarse-grain database $\DBCoarseEpoch$.
\end{itemize}
\textbf{Upon receiving a coarse-grain count query (CC)} (query point=$q$, radius=$a$) for epoch $\Epoch$ from a journalist:
\begin{itemize}[left=0.1cm]
\item Return the count of entries $\tilde{x}$ within $\DBCoarseEpoch$ that are within distance $a$ of $q$.
\end{itemize}
\textbf{Upon receiving a coarse-grain threshold query (CT)} (query point=$q$, radius=$a$, threshold $t$) for epoch $\Epoch$ from a journalist:
\begin{itemize}[left=0.1cm]
\item Return the threshold result, i.e., whether $\DBCoarseEpoch$ contains at least $t$ entries $\tilde{x}$ that are within distance $a$ of $q$, to the journalist.
\end{itemize}
\textbf{Upon receiving a fine-grain count query (FC)} (query point=$q$, radius=$a$) for epoch $\Epoch$ with budget $b$ from a journalist:
\begin{itemize}[left=0.1cm]
\item Check the current remaining budget $\EpsilonFEpoch$.  If it is less than $b$, return $\bot$ and exit.  Else, reduce $\EpsilonFEpoch$ by $b$.
\item Let $c$ be the count of $x$ in $\DBFineEpoch$ that are within distance $a$ of $q$.
\item Roll Laplace noise $s$ based on $b$ as  in \Cref{alg:query-fine-count}; send $(c+s)$ to the journalist.
\item If $\EpsilonFEpoch$ is now 0 (all budget is spent for this epoch), permanently delete $\DBFineEpoch$.  (All future queries for $\DBFineEpoch$ will be refused.)
\end{itemize}
\textbf{Upon receiving a fine-grain threshold query (FT)} (query point=$q$, radius=$a$, threshold=$t$) for epoch $\Epoch$ with budget $b$ from a journalist:
\begin{itemize}[left=0.1cm]
\item Check the current remaining budget $\EpsilonFEpoch$.  If it is less than $b$, return $\bot$ and exit.
\item Let $c$ be the count of $x$ in $\DBFineEpoch$ that are within distance $a$ of $q$.
\item Roll Laplace noise $u$ and $v$ based on $b$ as described in \Cref{alg:query-fine-threshold}.
\item Return to the journalist 1 if $c + v \ge t + u$, 0 otherwise (i.e. the noised threshold was met/exceeded by the noised count).
\item If the threshold was exceeded, lower $\EpsilonFEpoch$ by $b$ and inform the servers that the threshold was exceeded.
\item If $\EpsilonFEpoch$ is now 0 (all budget is spent for this epoch), permanently delete $\DBFineEpoch$. (All future queries for $\DBFineEpoch$ will be refused.)
\end{itemize}%
\end{minipage}%
} 
\caption{Ideal functionality for \Synopsis (simplified; see \Cref{alg:full-ideal-func} for the full version including MPC simulation details)}
\label{alg:ideal-func}
\label{fig:ideal-func}
\end{figure}
\begin{figure}
\fbox{\begin{minipage}{0.97\linewidth}
\footnotesize
\textbf{Algorithm}: Generation of fine-grained and coarse-grained databases 
\\
\textbf{Summary}: The final noised matrix $\tilde{X} = MP + \Delta$ will be $(\epsilon_J, \delta_J)$-differentially private as per Alg.\ 1 of Kenthapadi et al \cite{Kenthapadi_2013}.  We discuss generating and storing the $x$ and $\tilde{x}$ values in \Cref{alg:storage}.  We process the $x$ vectors the fine-grained queries, and the $\tilde{x}$ vectors in the coarse-grained queries.
\\
\textbf{Projection matrix P}: $\Projection$ is a $\EmbLength \times \JLTk$ matrix in which each element is sampled from  a Gaussian distribution $\mathcal{N}(0, {\sigma_\Projection^2})$, where $\sigma_\Projection = 1/\sqrt{\JLTk}.$
\\
\textbf{Gaussian noise matrix $\Delta$}:
$\Delta$ is a $\EmbDimInitial \times \EmbDimFinal$ matrix in which each element is sampled from $\mathcal{N}(0, \sigma_\Delta^2)$, where $\sigma_\Delta$ is given as a function of $\epsilon_J$, $\delta_J$, and $P$ in \Cref{eqn:noise-matrix-sigma}.
\\
\textbf{Inputs:}
\begin{itemize}[left=0.1cm]
\item Each row of $M$ is one un-perturbed $\EmbDimInitial$-vector embedding $x'$ built from a message.
\end{itemize}
\textbf{Outputs:}
\begin{itemize}[left=0.1cm]
\item Each row of $X = MP$ is one un-perturbed $\EmbDimFinal$-length vector $x$ = $x'P$.
\item Each row of $\tilde{X} = X \Delta = MP + \Delta$ is one perturbed $\EmbDimFinal$-length vector $\tilde{x}$.
\end{itemize}
\end{minipage}}
\caption{Algorithm: Database Generation}
\label{alg:noise-generation}
\label{alg:generatenoise}
\label{alg:kkmm}
\vspace{-0.5em}
\end{figure}

\begin{figure}[h]
\fbox{\begin{minipage}{0.97\linewidth}
\footnotesize
\textbf{Algorithm}: Fine-grained count query
\\
\textbf{Summary}: Within a particular epoch $\Epoch$, querier learns a noisy count of elements matching query $q$ (i.e. the number of $x$ points within L2 distance $a$ of $q$).
\\
\textbf{DP Mechanism}: Laplace mechanism (\cite{dwork_algorithmic_2013} Defn.\ 3.3).
\\
\textbf{Parameters}: Privacy parameter $\EpsilonFineCount$, remaining epoch budget $\EpsilonFEpoch$
\\
\textbf{Inputs}: 
\begin{itemize}[left=0.1cm]
    \item\textbf{Servers}: Stored $[x]$ and $[x^2]$ shares
    \item\textbf{Querier}: Query point $q$, radius $a$ (both public)
\end{itemize}
\textbf{Outputs}: 
\begin{itemize}[left=0.1cm]
    \item\textbf{Servers}: No output
    \item\textbf{Querier}: Count of points $x$ s.t. $\Dist(x, q) < a$, plus $\Lap(1/\EpsilonFineCount)$ noise.
\end{itemize}
\textbf{Algorithm}:
\begin{enumerate}[left=0.1cm]
    \item \label{alg:fc-start1} Server pre-processing: 
    \begin{enumerate}[left=0.1cm]
        \item Servers collectively generate shares of a sample $s$ from $\Lap(1/\EpsilonFineCount)$
        \item \label{alg:fc-start2} Generate multiplication triples to perform $N_C$ comparison operations.
    \end{enumerate}
    \item Queriers send $q$ and $a$ to all MPC servers.
    \item If $\EpsilonFineCount$ exceeds the remaining budget, servers return $\bot$.
    \item \label{alg:fc-d} For each point $x$, servers compute $[d] = [x^2] - 2q[x] + q^2$ (where the $q[x]$ and $q^2$ multiplications are scalar multiplications done elementwise).  This is the squared L2 distance between $x$ and $q$.
    \item \label{alg:fc-b} For each element, servers burn the necessary multiplication triples to compute $[b] = [(d < a^2)]$, a share of $b$ which is 1 if $d$ is less than $a^2$ (which is true if and only if the distance between $x$ and $q$ is less than $a$), 0 otherwise.
    \item \label{alg:fc-cprime} Over all $n$ elements $x$, compute $[c'] = \sum_{i=1}^n [x_i]$.
    \item \label{alg:fc-c} Servers compute $[c] = [c'] + [s]$ (adding the Laplace noise computed earlier).
    \item Servers send $[c]$ to the querier and reduce the remaining budget $\EpsilonFEpoch$ by $\EpsilonFineCount$.
    \item Querier reconstructs and outputs $c$, the noisy count of matches.
\end{enumerate}
\textbf{Privacy budget burned}: $\EpsilonFineCount$
\end{minipage}}
\caption{Algorithm: \QueryFineCount}
\label{alg:fc}
\label{alg:query-fine-count}
\end{figure}
\begin{figure}[h]
\fbox{\begin{minipage}{0.97\linewidth}
\footnotesize
\textbf{Algorithm}: Fine-grained threshold query
\\
\textbf{Summary}: Within a particular epoch, and for given threshold $t$, querier learns a bit representing whether the noisy count of messages matching a query is above or below (a noised version of) $t$ 
\\
\textbf{DP Mechanism}: Sparse Vector Mechanism (\cite{smith_svt} Alg.\ 2)
\\
\textbf{Parameters}: Privacy parameter $\EpsilonFineThreshold$, current remaining privacy budget $\EpsilonFEpoch$
\\
\textbf{Inputs}: 
\begin{itemize}[left=0.1cm]
    \item\textbf{Servers}: Stored $[x]$ and $[x^2]$ shares
    \item\textbf{Querier}: Query point $q$, radius $a$, threshold $t$ (all public)
\end{itemize}
\textbf{Outputs}: 
\begin{itemize}[left=0.1cm]
    \item\textbf{Servers}: No output
    \item\textbf{Querier}: 1 if noisy count of points $x$ s.t. $\Dist(x, q) < a$ is above $t$, else 0
\end{itemize}
\textbf{Algorithm}:
\begin{enumerate}[left=0.1cm]
    \item Server preprocessing:
    \begin{enumerate}[left=0.1cm]
        \item Servers jointly sample $[u]$ where $u \sim \Lap(2/\epsilon_T)$, for $N_T$ operations.
        \item Servers jointly sample $[v]$ where $v \sim \Lap(4/\epsilon_T)$, for $N_T$ operations.
        \item Servers jointly compute multiplication triples for $N_T$ queries.
        \item Servers initialize an empty dictionary of open queries.
    \end{enumerate}
    \item Follow steps \ref{alg:fc-start2}-\ref{alg:fc-cprime} of \QueryFineCount so that server has $[c']$, the exact count of messages that match query $q$.
    \item If $(q, a, t)$ is \emph{not} a key in the dictionary of open queries (with some $\hat{t}$), set $[\hat{t}] = t + [u]$ (noise the threshold), and add $((q,a,t), [\hat{t}])$ to the list of open queries (each server stores its share of $\hat{t}$).
    Else (if $(q,a,t)$ \emph{is} a key on the list of open queries), $[\hat{t}]$ is set as the previously stored value for $(q,a,t)$ (each server sets its share to what it had previously stored).
    \item Servers compute $[\hat{c}'] = [c'] + [v]$.  (Noise the count with Laplace noise.)
    \item Servers compute $[\tau] = [\hat{c}' \ge \hat{t}]$, shares of a Boolean value which is 1 if the (noised) result is above the (noised) threshold, 0 otherwise.
    \item Servers send $[\tau]$ to querier and to each other.
    \item Servers reconstruct $\tau$.
    If $\tau=1$, lower the remaining budget $\EpsilonFEpoch$ by $\EpsilonFineThreshold$, and servers remove $(q,a,t)$ and corresponding $[\hat{t}]$ shares from the dictionary of open queries.
    \item Querier reconstructs and outputs $\tau$
\end{enumerate}
\textbf{Privacy budget burned}: $\EpsilonFineThreshold$ if $\hat{c}' \ge \hat{t}$ (threshold was passed), else 0.
\end{minipage}}
\caption{Algorithm: \QueryFineThreshold}
\label{alg:ft}
\label{alg:query-fine-threshold}
\end{figure}
 
\subsection{Party and query summary}

\Synopsis's technical implementation is a multi-party computation (MPC) protocol among several parties (described in \Cref{sec:stakeholder-analysis}). 
\begin{itemize}
\item \textbf{Donors}, MPC input parties who contribute shares of message embedding vectors $x$ and perturbed vectors $\tilde{x}$.
\item \textbf{Servers}, run by the data stewards from \Cref{sec:stakeholder-analysis}, who are MPC compute parties with no inputs or outputs (aside from privacy budget tracking, public from an MPC perspective); they store the shares of the donated $x$ and $\tilde{x}$ vectors.
\item \textbf{Queriers}, journalists from \Cref{sec:stakeholder-analysis} are MPC output parties.  They also provide queries (known to Servers; \Cref{sec:query-privacy})
\end{itemize}

All query types follow similar logic:
When making a query, the querier (journalist) designates a set of epochs over which they would like to search, specifies the query type, and provides a \emph{query message vector} $q$, a \emph{radius} $\BallRadius$, and for threshold queries, a threshold $t$.
These are illustrated visually in \Cref{fig:enter-label-fg,fig:enter-label-cg}.
These query parameters are sent to the MPC servers run by the data stewards.

The MPC servers are storing shares of embedding vectors corresponding to messages that were given to them by donors: $x$ for the fine-grained embedding, and $\tilde{x}$ for the coarse grained embedding corresponding to a donated message.
Within each relevant epoch, the servers calculate under MPC the cosine distance between $q$ and each message embedding in the epoch, $x$ (for fine grained; resp.\ $\tilde{x}$ for coarse-grained). That distance is compared to $\BallRadius$; if the distance between $x$ and $q$ is at most $\BallRadius$, $x$ is a \emph{match}.

The servers then calculate the number of matches for $q$ during the relevant epochs, all still under MPC, yielding secret shares of the resulting number.  Those shares are post-processed depending on which query type was selected (always resulting in a DP query response) and then sent to the journalist, who reconstructs the response to their query. We note that these parameters, and those in \Cref{tab:variables}, are from the perspective of the technical architecture, not the query interface shown to journalists.

We provide a brief overview of our four query types.  The ideal functionality is given in \Cref{alg:ideal-func} with more details in \Cref{sec:all_queries}. 

\begin{itemize}
    \item \QueryFineCount (\textbf{FC}). Shown in \Cref{alg:fc} and as the red solid line of \Cref{fig:trendlines}. 
    Servers jointly compute the count of fine-grain messages matching the query.  The count is input to the Laplace Mechanism (\Cref{def:laplace}) and shares of the result are sent to the journalist.  

    This is the highest-accuracy but highest-cost query, spending $\EpsilonCountFine$ privacy budget.
    \item \QueryFineThreshold (\textbf{FT}). Shown in \Cref{alg:ft}. 
    Servers jointly compute the count of fine-grain messages matching the query. The count is input to the Sparse Vector Mechanism (\Cref{def:svt}) against the threshold $t$, outputting shares of True or False. 
    If True, this spends $\EpsilonFineThreshold$, else it spends 0.
    
    \item \QueryCoarseCount (\textbf{CC}). Shown in \Cref{alg:cc} and as the blue dotted line of \Cref{fig:trendlines}. 
    When intaking texts, the donors run the PrivateProjection local DP mechanism (\Cref{def:privateprojection}) on input vectors $x$ to get perturbed vectors $\tilde{x}$ (incurring a one-time cost of $\EpsilonCoarse$). Servers jointly compute the count of coarse-grained matches, the resulting shares are sent to the journalist.
    This is lower-accuracy than \textbf{FC} but costs 0.
    \item \QueryCoarseThreshold (\textbf{CT}). Shown in \Cref{alg:ct}. Like \textbf{CC}, but adds an additional comparison; journalists learn whether the threshold was crossed instead of the full count. This is less accurate than \textbf{FT} but costs no privacy budget.
\end{itemize}

\subsection{MPC malicious security and proofs}
Our main security guarantee is malicious security against any coalition of servers and queriers, as long as one server remains uncorrupted.
The full ideal functionality is shown in \Cref{alg:full-ideal-func}, and a simplified version that ignores the MPC simulation pieces is in \Cref{alg:ideal-func}.  The real protocols are shown in \Cref{alg:generatenoise,alg:query-fine-count,alg:query-fine-threshold,alg:storage,alg:query-coarse-count,alg:query-coarse-threshold}.

Because our entire query infrastructure is implemented in the generic malicious-secure MPC library MP-SPDZ \cite{mp-spdz}, we get our proof of security ``for free'' and do not need to write a custom proof for this specific algorithm being run under generic MPC.  (Correctness of the real protocols matching the functionality can be checked straightforwardly.)

The intuition of the confidentiality provided by the ideal functionality is that it answers the journalists' DP queries directly (with counts, not messages) and provides only random ``simulated'' shares with no relation to the inputs to the servers. As such the servers and queriers never have message information.  Server deviation from the protocol can be detected by having the ideal functionality ``check the servers' work'' by honestly running the algorithm on their messages; the servers themselves have no state secret from the ideal functionality so checking their work is trivial.

\subsection{Data Storage and Queries}
\label{sec:all_queries}
\subsubsection{Pre-processing}
\label{sec:projection}

The storage and pre-query processing of input messages, which happens at intake before any of these queries are made, is contained in \Cref{alg:storage} in \Cref{app:extra-algs}. This describes the process of donors taking their message $m$, computing a BERT embedding $x'$ of that message, using the projection matrix $P$ described in \Cref{alg:generatenoise} to project $x'$ down to a smaller vector $x$, and then adding Gaussian noise to yield the perturbed vector $\tilde{x}$.
In addition to storing shares $[x]$ and $[\tilde{x}]$, the servers also store elementwise squares $[x^2]$ and $[\tilde{x}^2]$ to reduce online computation time.

We highlight the dimension reduction step.
Using a random projection improves both latency and, as discussed in \Cref{sec:coarse_grained}, also reduces the amount of noise needed for coarse-grained queries.

\Cref{alg:kkmm} shows the DP PrivateProjection mechanism of Kenthapadi et al.\ \cite{Kenthapadi_2013}, 
which uses a dimension-reducing projection based on the Johnson-Lindenstrauss (JLT or JL) \cite{lindenstrauss1984extensions} followed by noise addition.
The JLT preserves pairwise L2 distances (therefore also $\Dist$ for normed vectors, see \Cref{app:cosine-facts}) within a factor of $\JLParam$ w.h.p.:
\begin{align*}
(1 - \JLParam) \Vert x - y \Vert^2 \le \Vert xP - yP \Vert^2 \le (1 + \JLParam) \Vert x - y \Vert^2.
\end{align*}

Parameter $\JLParam$ affects the final dimension $\JLTk$; as in  \cite{Kenthapadi_2013}, we require $\JLTk = \Omega((\log n)/\JLParam^2)$.
See \Cref{tab:proj_benchmarks} for benchmarks for $\JLParam$.
In \Cref{sec:reuse-jlt} we discuss how we ``reuse'' the JLT for both efficiency and accuracy purposes in the coarse-grained queries.
\subsubsection{Fine-grained queries}
\label{sec:fine_grained}

\textbf{FC} (\Cref{alg:fc}).  
\QueryFineCount queries (\textbf{FC}) are targeted queries with a high accuracy and a corresponding privacy budget spend per-query, $\EpsilonFineCount$.  They return the Laplace mechanism (\cite{dwork_algorithmic_2013} Defn.\ 3.3) applied to counts of matching queries; the full algorithm is given in \Cref{alg:fc}.  In essence, the MPC servers compute a secret-shared bit per each message that represents whether its distance is within \BallRadius of the query. those bits are then summed to obtain the count of matches, Laplace noise is added under MPC (computed during preprocessing).
FC queries are suitable for close inspection of topics that are already known to be of interest (e.g. surfaced by a different query or known from external factors).

The \QueryFineThreshold (\textbf{FT}; \Cref{alg:ft}) algorithm
is based on the Sparse Vector Mechanism  (\cite{smith_svt} Alg.\ 2, a variant on \cite{DBLP:conf/stoc/DworkNRRV09}), which allows repeated queries to check whether a counts are above or below a noised threshold $t$.  
This mechanism's privacy budget $\EpsilonFineThreshold$ is burned the first time a query's (noised) count result surpasses the (also noised) threshold, but not until then as long as all queries remain below the threshold.
The servers also learn the single bit of whether the threshold was met, as this is needed to appropriately track the spent budget.
FT queries can be a way to track a topic that is expected to be low until an event occurs.  They are more accurate than coarse-grained queries, and cheaper than FC queries (if the threshold is not exceeded immediately).

We note that since queries are known to the servers, fine-grained query results per epoch can be cached by all servers to avoid over-spending.
When the budget for an epoch is depleted, the fine grained data is deleted, leaving only coarse-grained for that epoch.
\subsubsection{Coarse-grained queries}
\label{sec:coarse_grained}
\label{sec:coarse-grained}
\label{sec:coarse-grain}
Coarse-grained queries (\textbf{CC}, \textbf{CT}) differ from fine-grained because they directly perturb message embeddings using a variant of the Gaussian mechanism (\Cref{def:gaussian}).  With an up-front privacy cost $\EpsilonCoarse$, this allows unlimited querying of the perturbed points while spending no further budget due to post-processing properties of DP. The tradeoff is reduced accuracy.  

\label{sec:cc-overestimate}
This inaccuracy is not merely due to noise greater than that of the fine-grained regime; it stems from two factors.
First, since a count of perturbed matches can never be below zero, there will be a natural skew up for low-true-count queries (e.g. if the count of true matches is zero, the count of perturbed points cannot be lower).
The second factor depends on the data around the query.  For a query within a large cluster, more nearby neighboring vectors will likely be pushed into the query than matching vectors will be pushed out, causing an overestimate (indeed, this is likely what we see in \Cref{fig:trendlines}).
On the other hand, if the query captures all points in a sparse space, more vectors will likely be pushed out of the region than added in.

This observation is a general property of DP mechanisms that perturb data input to a function, rather than altering outputs of a function, one that is not captured by mere differences in epsilon.

Despite the inaccuracy, coarse-grained queries offer the major benefit of reusable zero-cost querying.
All privacy cost is paid when intaking the message.
The \textbf{CC} (\Cref{alg:cc}) and \textbf{CT}  (\Cref{alg:ct}) queries are straightforward MPC count and thresholding algorithms (\Cref{app:coarse-figures}).

\textbf{Reusing dimension reduction to reduce added noise.}
\label{sec:reuse-jlt}
One way we could have created the perturbed version of our dataset would have been to directly use the Gaussian mechanism \cite{DBLP:conf/eurocrypt/DworkKMMN06,dwork_algorithmic_2013}.
However, Kenthapadi et al.\ \cite{Kenthapadi_2013} show that performing a random projection $P$ (see \Cref{alg:kkmm}) and then adding Gaussian noise afterward allows us to reduce the amount of noise added.
Therefore, we ``reuse'' the dimension reduction by utilizing it as part of the DP PrivateProjection mechanism of Kenthapadi et al, saving a factor of approximately $(n/k)^2$ in the variance of the Gaussian noise used to perturb the data. \cite{Kenthapadi_2013}. Vectors generated by our BERT embedding model are normalized to unit length 1 by the embedding model; we re-normalize these vectors after projection.

The additive noise $\NoiseMatrix$ is randomly sampled from Gaussian normal distribution $\mathcal{\NoiseMatrix}(0, \sigma_\NoiseMatrix^2),$  where 
\begin{equation}
\sigma_\NoiseMatrix \ge \omega_2(P)\frac{\sqrt{2(\ln(\frac{1}{2\delta})+\epsilon_J)}}{\epsilon_J}, 
\label{eqn:noise-matrix-sigma}
\end{equation}
where $\omega_2(\Projection) = \max_{e_i} \Vert e_i P \Vert_2$ is the greatest L2 distance between two embedding vectors; $P$ was chosen such that expected $\ell_2$ sensitivity is tightly concentrated around 1 \cite{kasiviswanathan_ldp_2008}.
\section{Implementation and Evaluation}
\label{sec:implementation}

This section contains our implementation, testing, and deployment of \Synopsis in the setting of a real-world investigation: we benchmark and test \Synopsis on a corpus of 34K Hindi messages extracted from a dozen WhatsApp groups with mostly India-based members discussing Indian politics.

Additional information about our dataset follows in the next subsection (\Cref{sec:investigation}). We perform our latency tests on an abridged and full corpus of 3,442 and 34,024 messages, respectively, drawn from Hindi WhatsApp public groups.

\subsection{WhatsApp discussions about Ram Mandir}
\label{sec:investigation}
\label{sec:example}
\label{sec:impl_dataset}

\textit{Ram Mandir,} known also as the Ram Temple, is a temple complex in Ayodhya, India. Online discussion of Ram's construction often occurs in close association with heated debates about national politics. We worked with journalists
to conduct post hoc analyses of a dozen WhatsApp groups whose India-based members frequently discuss national politics.
Using a Ram-related message as our query, we were able to trace the rise and fall in frequency of Ram-related discussion across all groups, shown in \Cref{fig:trendlines} on \cpageref{fig:trendlines}.
We first use \textbf{CC} queries to understand baseline levels of message frequency about the Ram Temple, examining messages sent over the course of 410 days. After identifying peaks in our coarse-grained trend line around the middle and end of January 2024, we proceeded \textbf{FC} queries which correctly identified the day of the temple's inauguration: January 22, 2024.

\textbf{Datasets.} The full dataset of 34,024 WhatsApp messages we used for our benchmarks was extracted from 12 public WhatsApp groups. The earliest timestamp in the full message corpus is June 14, 2023 and the most recent message in the corpus dates from November 4, 2024. The abridged dataset we use for our accuracy tests and inauguration day analysis comprises 3,442 message embeddings dating from January 9, 2024 through February 3, 2024. 

\subsection{Implementation setup}
\label{sec:general_impl}
\label{sec:impl_setup}
Each regime is implemented in MP-SPDZ \cite{mp-spdz}, a popular MPC library. We conduct all benchmark tests on a machine with an AMD EPYC 7742 64-Core Processor with 1007 gigabytes of memory. Results reported in this section utilized 64 threads.
In the exemplar analysis presented in this section, all fine- and coarse-grained queries are conducted over 500-dimension embedding vectors. 

\subsection{Benchmark measurements and results.}
\label{sec:impl_results}

We test query accuracy and latency on the dataset discussed in \Cref{sec:impl_dataset} and on the setup described in \Cref{sec:impl_setup}.

\textbf{Latency.} As shown in \Cref{tab:proj_benchmarks,tab:mega-table}, we achieve roughly 1 ms latency per-database element (1 $\JLTk$-length vector) for all query types, with overall run time scaling linearly with the total size of the database(s) submitted by donors. 

Queries issued to a database comprising 3,442 message embeddings, each of dimension 300, required about 3 seconds to complete, averaged over 1000 queries.

\begin{table}[!t]
\centering
\def\arraystretch{0.9}
\caption{Weighted average error for settings of $\JLTk$. ``Error (/ day)'' is weighted by each epoch's contribution to the cumulative match count.  ``Error (cumul.)'' is the sum total of all matches, observed across all epochs (days), divided by the ground truth total match count (75). Latency is per database element.
}
    \begin{tabular}{c|c|c|c|c}
        $\JLParam$ & $\JLTk$ & Error (/ epoch) & Error (cumul.) & Time/elem (ms), $\sigma$ \\ 
        \toprule
        \textbf{0.4} & 555 & 0.05 & 0.013 & 1.1, 0.35 \\ 
        \midrule
        \textbf{0.5} & 390 & 0.12 & 0.04 & 1.0, 0.35 \\ 
        \midrule
        \textbf{0.6} & 300 & 0.12 & 0.04 & 0.9, 0.4 \\ 
        \midrule
        \textbf{0.7} & 249 & 0.23 & 0.067 & 0.8, 0.32 \\ 
        \midrule
        \textbf{0.8} & 218 & 0.28 & 0.09 & 0.7, 0.29 \\ 
        \midrule
    \end{tabular}
    \label{tab:proj_benchmarks}
\end{table}

\def\themult{0.1}
\begin{table}
\caption{Error for all queries.  Weighted average error (MAE) for counts, true and false positive rates (TPR/FPR) for thresholds.  Latency is per database element. $\JLTk=500$.}
\label{tab:mega-table}
\label{tab:mega-tabular}
\vspace{-0.2cm}
\resizebox{\linewidth}{!}{
\begin{tabular}{cc}
\begin{tabular}{p{0.05\linewidth}p{\themult\linewidth}p{\themult\linewidth}p{\themult\linewidth}}
        & \rot{$\EpsilonCountFine$ | $\EpsilonCountCoarse$} & \rot{Error (MAE)} & \rot{Time/elem (ms), $\sigma$} \\ 
        \toprule
        \multirow{8}{*}{\textbf{FC}} 
                & 0.2 & 4.388 & 1.0, 0.4 \\ 
                & 0.4 & 3.472 & 1.0, 0.3 \\
                & 0.6 & 1.345 & 1.0, 0.3 \\
                & 0.8 & 0.786 & 1.1, 0.4 \\
                & 1   & 0.612 & 1.0, 0.4 \\
                & 2   & 0.455 & 1.1, 0.3 \\
                & 3   & 0.320 & 1.0, 0.3 \\
                & 4   & 0.140 & 1.1, 0.3 \\ 
                \midrule
        \multirow{8}{*}{\textbf{CC}} 
                & 0.2 & 65      & 1.0,  1.8 \\ 
                & 0.4 & 65.38   & 1.0, 2.0 \\
                & 0.6 & 54.85   & 1.0, 2.2 \\
                & 0.8 & 44.96   & 1.0, 2.1 \\
                & 1   & 38.61   & 1.0, 2.2 \\
                & 2   & 14      & 1.0, 2.6 \\
                & 3   &  3      & 1.1, 2.5 \\
                & 4   &  2.192  & 1.1, 2.0 \\ 
                \midrule
    \end{tabular}
&
    \begin{tabular}{p{0.05\linewidth}p{\themult\linewidth}p{\themult\linewidth}p{\themult\linewidth}p{\themult\linewidth}}
        & \rot{$\EpsilonFineThreshold$ | $\EpsilonCoarse$} & \rot{TPR} & \rot{FPR} & \rot{Time/elem (ms), $\sigma$} \\ 
        \toprule
        \multirow{8}{*}{\textbf{FT}} 
                & 0.2   &  3/3  &  0/23  & 1.1, 0.3 \\ 
                & 0.4   &  3/3  &  0/23  & 1.0, 0.3 \\
                & 0.6   &  3/3  &  0/23  & 1.0, 0.4 \\
                & 0.8   &  3/3  &  0/23  & 1.0, 0.3 \\
                & 1     &  3/3  &  0/23  & 1.0, 0.3 \\
                & 2     &  3/3  &  0/23  & 1.0, 0.4 \\
                & 3     &  3/3  &  0/23  & 1.0, 0.3 \\
                & 4     &  3/3  &  0/23  & 1.0, 0.2 \\
                \midrule
        \multirow{8}{*}{\textbf{CT}}
                & 0.2 &  1/1  & 10/27 & 1.0, 1.3 \\ 
                & 0.4 &  1/1  & 9/26 & 1.0, 1.5 \\
                & 0.6 &  1/1  & 8/25 & 1.1, 1.7\\
                & 0.8 &  1/1  & 7/24 & 0.9, 1.8  \\
                & 1   &  1/1  & 7/24 & 1.1, 2.0 \\
                & 2   &  1/1  & 6/23 & 1.0, 1.3 \\
                & 3   &  1/1  & 5/22 & 1.1, 2.0 \\
                & 4   &  1/1  & 5/22  & 1.0, 1.6 \\ 
                \midrule
    \end{tabular}
\end{tabular}
} 
\end{table}

\textbf{Accuracy.} 
In\ \Cref{tab:proj_benchmarks,tab:mega-table}, we report two types of error rates: a weighted average of daily counts results, wherein each error contribution is weighted by the proportion of the whole corpus's counts contributed by that epoch's messages (``Error (/ epoch)''); and percent error with respect to the total number of counts observed across the entire corpus without per-epoch (per-day) buckets (``Error (cumul.)''). These two metrics yield complementary insights about our query regimes: our per-day error rates track the absolute magnitude of noise injected point-wise, per epoch. Our cumulative error rates, which are generally lower than our weighted per-day error rates, indicate that these noise levels average out, on balance, across multiple epochs of querying.

\textbf{Parameter tuning.} Accuracy can modestly improve with higher projection dimension (we test with 300-dimension transformed SBERT embeddings) and larger privacy budgets, as shown in \Cref{tab:proj_benchmarks,tab:mega-table}. Our distance calculations after projection and encryption have high accuracy---above 90\%---with respect to ground-truth distance measurements.
MPC does not introduce additional sources of error with our default secret float precision levels (16 bit floats, and 32 bits overall).
While bounded sum functions can underestimate in some DP libraries \cite{casacuberta2022widespread}, our use of these is limited (at most twice) and we are confident this error is negligible in our use case. 

\textbf{Summary.} In our experimental investigation, \SVS maintained reporting accuracy above 90\% for per-day and per-month queries, over fine-grained databases comprising 3,442 and 34,024 entries. We are able to verify the correctness of our investigation: the trends we identify align with offline conversations about the Ram Temple, the subject of our query.

\bibliographystyle{unsrt}
\bibliography{references}

\begin{thebibliography}{100}

\bibitem{cummings2024advancing}
Rachel Cummings, Damien Desfontaines, David Evans, Roxana Geambasu, Yangsibo Huang, Matthew Jagielski, Peter Kairouz, Gautam Kamath, Sewoong Oh, Olga Ohrimenko, et~al.
\newblock Advancing differential privacy: Where we are now and future directions for real-world deployment.
\newblock {\em Harvard data science review}, 6(1), 2024.

\bibitem{nanayakkara2024measure}
Priyanka Nanayakkara, Hyeok Kim, Yifan Wu, Ali Sarvghad, Narges Mahyar, Gerome Miklau, and Jessica Hullman.
\newblock Measure-observe-remeasure: An interactive paradigm for differentially-private exploratory analysis.
\newblock In {\em 2024 IEEE Symposium on Security and Privacy (SP)}, pages 1047--1064. IEEE, 2024.

\bibitem{statistaWhatsAppMarket}
Statista.
\newblock Whatsapp penetration rate among global messaging app users as of april 2022, by country, September 2023.

\bibitem{wa_about_e2ee}
{WhatsApp Help Center}.
\newblock About end-to-end encryption, 2024.
\newblock \url{https://faq.whatsapp.com/820124435853543}.

\bibitem{double_ratchet}
Moxie~Marlinspike Trevor~Perrin.
\newblock The double ratchet algorithm, 11 2016.
\newblock \url{https://signal.org/docs/specifications/doubleratchet/}.

\bibitem{domingo2023india}
Aldohn Domingo.
\newblock India cracks down on ai misinformation, whatsapp may be required to display first message source.
\newblock {\em Tech Times}.
\newblock \url{https://www.techtimes.com/articles/297566/20231016/india-ai-misinformation-whatsapp-first-source.htm}.

\bibitem{shih2023inside}
Gerry Shih.
\newblock Inside the vast digital campaign by hindu nationalists to inflame india.
\newblock {\em The Washington Post}.
\newblock \url{https://www.washingtonpost.com/world/2023/09/26/hindu-nationalist-social-media-hate-campaign/}.

\bibitem{lamb2024generative}
Kate Lamb, Fanny Potkin, and Ananda Teresia.
\newblock Generative ai may change elections this year. indonesia shows how.
\newblock {\em Reuters}.
\newblock \url{https://www.reuters.com/technology/generative-ai-faces-major-test-indonesia-holds-largest-election-since-boom-2024-02-08/}.

\bibitem{sekargati2024hate}
Jati~Savitri Sekargati.
\newblock Hate speech is likely to intensify on social media ahead of indonesia’s election.
\newblock {\em The Conversation}.
\newblock \url{https://theconversation.com/hate-speech-is-likely-to-intensify-on-social-media-ahead-of-indonesias-election-211896}.

\bibitem{bursztynThousandsOfSmall}
Victor~S. Bursztyn and Larry Birnbaum.
\newblock Thousands of small, constant rallies: a large-scale analysis of partisan whatsapp groups.
\newblock In {\em Proceedings of the 2019 IEEE/ACM International Conference on Advances in Social Networks Analysis and Mining}, ASONAM '19, page 484–488, New York, NY, USA, 2020. Association for Computing Machinery.

\bibitem{de2021digital}
Ivandro~Claudino de~S{\'a}, Jos{\'e}~Maria Monteiro, Jos{\'e} Wellington~Franco da~Silva, Leonardo~Monteiro Medeiros, Pedro Jorge~Chaves Mourao, and Lucas Cabral~Carneiro da~Cunha.
\newblock Digital lighthouse: A platform for monitoring public groups in whatsapp.
\newblock In {\em ICEIS (1)}, pages 297--304, 2021.

\bibitem{agarwalJettisoningJunk}
Pushkal Agarwal, Aravindh Raman, Damiola Ibosiola, Nishanth Sastry, Gareth Tyson, and Kiran Garimella.
\newblock Jettisoning junk messaging in the era of end-to-end encryption: A case study of whatsapp.
\newblock In {\em Proceedings of the ACM Web Conference 2022}, WWW '22, page 2582–2591, New York, NY, USA, 2022. Association for Computing Machinery.

\bibitem{rosenfeld2018studywhatsappusagepatterns}
Avi Rosenfeld, Sigal Sina, David Sarne, Or~Avidov, and Sarit Kraus.
\newblock A study of whatsapp usage patterns and prediction models without message content, 2018.
\newblock \url{https://arxiv.org/abs/1802.03393}.

\bibitem{seufert2015analysis}
Michael Seufert, Anika Schwind, Tobias Ho{\ss}feld, and Phuoc Tran-Gia.
\newblock Analysis of group-based communication in whatsapp.
\newblock In {\em Mobile Networks and Management: 7th International Conference, MONAMI 2015, Santander, Spain, September 16-18, 2015, Revised Selected Papers 7}, pages 225--238. Springer, 2015.

\bibitem{kazemi2022a}
A~Kazemi, K~Garimella, GK~Shahi, D~Gaffney, and SA~Hale.
\newblock Research note: Tiplines to uncover misinformation on encrypted platforms: a case study of the 2019 indian general election on whatsapp.
\newblock {\em Harvard Kennedy School Misinformation Review}, 3(1), 2022.

\bibitem{arun2019whatsapp}
Chinmayi Arun.
\newblock On whatsapp, rumours, lynchings, and the indian government.
\newblock {\em Economic \& Political Weekly}, 54(6), 2019.

\bibitem{themarkupBuiltFacebook}
Surya Mattu, Leon Yin, Angie Waller, and Jon Keegan.
\newblock {H}ow {W}e {B}uilt a {F}acebook {I}nspector – {T}he {M}arkup --- themarkup.org, 2021.
\newblock [Accessed 07-02-2024].

\bibitem{garimella2024whatsappexplorerdatadonation}
Kiran Garimella and Simon Chauchard.
\newblock Whatsapp explorer: A data donation tool to facilitate research on whatsapp, 2024.
\newblock \url{https://arxiv.org/abs/2404.01328}.

\bibitem{whatsapp_monitor}
Simon Chauchard and Kiran Garimella.
\newblock Collecting whatsapp data for social science research: Challenges and a proposed solution., 9 2023.
\newblock \url{https://gvrkiran.github.io/content/WhatsApp_Chauchard_Garimella_AUGUST2023_V3.pdf}.

\bibitem{nio}
David Lazer, David Choffnes, and Christopher Wilson.
\newblock National internet observatory, 2024.
\newblock https://nationalinternetobservatory.org.

\bibitem{feal2024introduction}
Alvaro Feal, Jeffrey Gleason, Pranav Goel, Jason Radford, Kai-Cheng Yang, John Basl, Michelle Meyer, David Choffnes, Christo Wilson, and David Lazer.
\newblock Introduction to national internet observatory.
\newblock 2024.

\bibitem{google_cdlp}
Scott Ellis.
\newblock Take charge of your sensitive data with the cloud data loss prevention (dlp) api, 3 2018.
\newblock \url{https://cloud.google.com/blog/products/gcp/take-charge-of-your-sensitive-data-with-the-cloud-dlp-api/}.

\bibitem{dwork_2006}
Cynthia Dwork, Frank McSherry, Kobbi Nissim, and Adam Smith.
\newblock Calibrating noise to sensitivity in private data analysis.
\newblock In {\em Proceedings of the Third Conference on Theory of Cryptography}, TCC'06, page 265–284, Berlin, Heidelberg, 2006. Springer-Verlag.

\bibitem{ohm2009broken}
Paul Ohm.
\newblock Broken promises of privacy: Responding to the surprising failure of anonymization.
\newblock {\em UCLA l. Rev.}, 57:1701, 2009.

\bibitem{song2020information}
Congzheng Song and Ananth Raghunathan.
\newblock Information leakage in embedding models.
\newblock In {\em Proceedings of the 2020 ACM SIGSAC conference on computer and communications security}, pages 377--390, 2020.

\bibitem{morris2023text}
John~X. Morris, Volodymyr Kuleshov, Vitaly Shmatikov, and Alexander~M. Rush.
\newblock Text embeddings reveal (almost) as much as text, 2023.

\bibitem{morris2023language}
John~X. Morris, Wenting Zhao, Justin~T. Chiu, Vitaly Shmatikov, and Alexander~M. Rush.
\newblock Language model inversion, 2023.

\bibitem{hitaj_deep_2017}
Briland Hitaj, Giuseppe Ateniese, and Fernando Perez-Cruz.
\newblock Deep models under the {GAN}: Information leakage from collaborative deep learning.
\newblock In {\em Proceedings of the 2017 {ACM} {SIGSAC} Conference on Computer and Communications Security}, pages 603--618. {ACM}.

\bibitem{li_towards_2018}
Yitong Li, Timothy Baldwin, and Trevor Cohn.
\newblock Towards robust and privacy-preserving text representations.

\bibitem{gault_2025}
Researchers scrape 2 billion discord messages and publish them online.
\newblock May 20125.

\bibitem{aquino2025discord}
Yan Aquino, Pedro Bento, Arthur Buzelin, Lucas Dayrell, Samira Malaquias, Caio Santana, Victoria Estanislau, Pedro Dutenhefner, Guilherme~HG Evangelista, Luisa~G Porf{\'\i}rio, et~al.
\newblock Discord unveiled: A comprehensive dataset of public communication (2015-2024).
\newblock {\em arXiv preprint arXiv:2502.00627}, 2025.

\bibitem{shekhar_guns_2024}
Raj Shekhar.
\newblock Unveiling the dark trade: The rise of illicit weapons in india.
\newblock {\em The Times of India}, 2024.

\bibitem{ram_scam_2024}
TOI~Tech Desk.
\newblock Scam alert: These whatsapp messages on ayodhya ram mandir are fake and ‘dangerous’.
\newblock {\em Times of India}, January 2024.

\bibitem{avelar_bolsonaro_2019}
Daniel Avelar.
\newblock Whatsapp fake news during brazil election ‘favoured bolsonaro’.
\newblock {\em The Guardian}, October 2019.

\bibitem{jaswal_bjp_2024}
Srishti Jaswal.
\newblock Inside the bjp’s whatsapp machine.
\newblock {\em Rest of World}, May 2024.

\bibitem{mattu_bjp_2024}
Surya Mattu, Jeremy Singer-Vine, Qazi Firas, and Pallab Deb.
\newblock Making the political personal: How the bjp spread their message through whatsapp in an indian town.
\newblock {\em Digital Witness Lab Methodologies}, May 2024.

\bibitem{lankesh1}
Surya Mattu and Micha Gorelick.
\newblock A video helped incite the murder of a prominent bangalore journalist, we investigated its spread on social media.
\newblock {\em Digital Witness Lab}.
\newblock \url{https://www.digitalwitnesslab.org/outputs/a-video-helped-incite-the-murder-of-a-prominent-bangalore-journalist-we-investigated-its-spread-on-social-media}.

\bibitem{lankesh2}
Phineas Rueckert.
\newblock In the age of false news: A journalist, a murder, and the pursuit of an unfinished investigation in india.
\newblock {\em Forbidden Stories}.
\newblock \url{https://forbiddenstories.org/story-killers/gauri-lankesh-in-the-age-of-false-news/}.

\bibitem{sweeney1996replacing}
Latanya Sweeney.
\newblock Replacing personally-identifying information in medical records, the scrub system.
\newblock In {\em Proceedings of the AMIA annual fall symposium}, page 333. American Medical Informatics Association, 1996.

\bibitem{douglass2005identification}
MM~Douglass, GD~Cliffford, Andrew Reisner, WJ~Long, GB~Moody, and RG~Mark.
\newblock De-identification algorithm for free-text nursing notes.
\newblock In {\em Computers in Cardiology, 2005}, pages 331--334. IEEE, 2005.

\bibitem{deleger2013large}
Louise Deleger, Katalin Molnar, Guergana Savova, Fei Xia, Todd Lingren, Qi~Li, Keith Marsolo, Anil Jegga, Megan Kaiser, Laura Stoutenborough, et~al.
\newblock Large-scale evaluation of automated clinical note de-identification and its impact on information extraction.
\newblock {\em Journal of the American Medical Informatics Association}, 20(1):84--94, 2013.

\bibitem{dernoncourt2017identification}
Franck Dernoncourt, Ji~Young Lee, Ozlem Uzuner, and Peter Szolovits.
\newblock De-identification of patient notes with recurrent neural networks.
\newblock {\em Journal of the American Medical Informatics Association}, 24(3):596--606, 2017.

\bibitem{johnson2020deidentification}
Alistair~EW Johnson, Lucas Bulgarelli, and Tom~J Pollard.
\newblock Deidentification of free-text medical records using pre-trained bidirectional transformers.
\newblock In {\em Proceedings of the ACM Conference on Health, Inference, and Learning}, pages 214--221, 2020.

\bibitem{anandan2012t}
Balamurugan Anandan, Chris Clifton, Wei Jiang, Mummoorthy Murugesan, Pedro Pastrana-Camacho, and Luo Si.
\newblock t-plausibility: Generalizing words to desensitize text.
\newblock {\em Trans. Data Priv.}, 5(3):505--534, 2012.

\bibitem{chakaravarthy2008efficient}
Venkatesan~T Chakaravarthy, Himanshu Gupta, Prasan Roy, and Mukesh~K Mohania.
\newblock Efficient techniques for document sanitization.
\newblock In {\em Proceedings of the 17th ACM conference on Information and knowledge management}, pages 843--852, 2008.

\bibitem{sanchez2016c}
David S{\'a}nchez and Montserrat Batet.
\newblock C-sanitized: A privacy model for document redaction and sanitization.
\newblock {\em Journal of the Association for Information Science and Technology}, 67(1):148--163, 2016.

\bibitem{feyisetan_perturb_2020}
Oluwaseyi Feyisetan, Borja Balle, Thomas Drake, and Tom Diethe.
\newblock Privacy- and utility-preserving textual analysis via calibrated multivariate perturbations.
\newblock In {\em Proceedings of the 13th International Conference on Web Search and Data Mining}, WSDM '20, page 178–186, New York, NY, USA, 2020. Association for Computing Machinery.

\bibitem{xu_density-aware_2021}
Nan Xu, Oluwaseyi Feyisetan, Abhinav Aggarwal, Zekun Xu, and Nathanael Teissier.
\newblock Density-aware differentially private textual perturbations using truncated gumbel noise.
\newblock 34(1).

\bibitem{xu_utilitarian_2021}
Zekun Xu, Abhinav Aggarwal, Oluwaseyi Feyisetan, and Nathanael Teissier.
\newblock On a utilitarian approach to privacy preserving text generation.
\newblock In {\em Proceedings of the Third Workshop on Privacy in Natural Language Processing}, pages 11--20. Association for Computational Linguistics.

\bibitem{yue_differential_2021}
Xiang Yue, Minxin Du, Tianhao Wang, Yaliang Li, Huan Sun, and Sherman S.~M. Chow.
\newblock Differential privacy for text analytics via natural text sanitization.

\bibitem{qu_natural_2021}
Chen Qu, Weize Kong, Liu Yang, Mingyang Zhang, Michael Bendersky, and Marc Najork.
\newblock Natural language understanding with privacy-preserving {BERT}.
\newblock In {\em Proceedings of the 30th {ACM} International Conference on Information \& Knowledge Management}, pages 1488--1497. {ACM}.

\bibitem{chen_customized_2023}
Sai Chen, Fengran Mo, Yanhao Wang, Cen Chen, Jian-Yun Nie, Chengyu Wang, and Jamie Cui.
\newblock A customized text sanitization mechanism with differential privacy.
\newblock In {\em Findings of the Association for Computational Linguistics: {ACL} 2023}, pages 5747--5758. Association for Computational Linguistics.

\bibitem{meehan_sentence-level_2022}
Casey Meehan, Khalil Mrini, and Kamalika Chaudhuri.
\newblock Sentence-level privacy for document embeddings.

\bibitem{weggenmann_dp-vae_2022}
Benjamin Weggenmann, Valentin Rublack, Michael Andrejczuk, Justus Mattern, and Florian Kerschbaum.
\newblock {DP}-{VAE}: Human-readable text anonymization for online reviews with differentially private variational autoencoders.
\newblock In {\em Proceedings of the {ACM} Web Conference 2022}, pages 721--731. {ACM}.

\bibitem{krishna_adept_2021}
Satyapriya Krishna, Rahul Gupta, and Christophe Dupuy.
\newblock {ADePT}: Auto-encoder based differentially private text transformation.
\newblock In {\em Proceedings of the 16th Conference of the European Chapter of the Association for Computational Linguistics: Main Volume}, pages 2435--2439. Association for Computational Linguistics.

\bibitem{habernal_when_2021}
Ivan Habernal.
\newblock When differential privacy meets {NLP}: The devil is in the detail.
\newblock In {\em Proceedings of the 2021 Conference on Empirical Methods in Natural Language Processing}, pages 1522--1528. Association for Computational Linguistics.

\bibitem{mattern_limits_2022}
Justus Mattern, Benjamin Weggenmann, and Florian Kerschbaum.
\newblock The limits of word level differential privacy.

\bibitem{DBLP:journals/siamcomp/KasiviswanathanLNRS11}
Shiva~Prasad Kasiviswanathan, Homin~K. Lee, Kobbi Nissim, Sofya Raskhodnikova, and Adam~D. Smith.
\newblock What can we learn privately?
\newblock {\em {SIAM} J. Comput.}, 40(3):793--826, 2011.

\bibitem{mp-spdz}
Marcel Keller.
\newblock {MP-SPDZ}: A versatile framework for multi-party computation.
\newblock In {\em Proceedings of the 2020 ACM SIGSAC Conference on Computer and Communications Security}, 2020.

\bibitem{melis_efficient}
Luca Melis, George Danezis, and Emiliano De~Cristofaro.
\newblock Efficient private statistics with succinct sketches.
\newblock {\em arXiv preprint arXiv:1508.06110}, 2015.

\bibitem{hagen_p2kmv}
Hagen Sparka, Florian Tschorsch, and Bj{\"o}rn Scheuermann.
\newblock P2kmv: A privacy-preserving counting sketch for efficient and accurate set intersection cardinality estimations.
\newblock {\em Cryptology ePrint Archive}, 2018.

\bibitem{boneh_lightweight}
Dan Boneh, Elette Boyle, Henry Corrigan-Gibbs, Niv Gilboa, and Yuval Ishai.
\newblock Lightweight techniques for private heavy hitters.
\newblock In {\em 2021 IEEE Symposium on Security and Privacy (SP)}, pages 762--776. IEEE, 2021.

\bibitem{whisper}
Mayank Rathee, Yuwen Zhang, Henry Corrigan{-}Gibbs, and Raluca~Ada Popa.
\newblock Private analytics via streaming, sketching, and silently verifiable proofs.
\newblock In {\em {IEEE} Symposium on Security and Privacy, {SP} 2024, San Francisco, CA, USA, May 19-23, 2024}, pages 3072--3090. {IEEE}, 2024.
\newblock \url{https://doi.org/10.1109/SP54263.2024.00245}.

\bibitem{compass}
Jinhao Zhu, Liana Patel, Matei Zaharia, and Raluca~Ada Popa.
\newblock Compass: Encrypted semantic search with high accuracy.
\newblock {\em {IACR} Cryptol. ePrint Arch.}, page 1255, 2024.
\newblock \url{https://eprint.iacr.org/2024/1255}.

\bibitem{tiptoe}
Alexandra Henzinger, Emma Dauterman, Henry Corrigan-Gibbs, and Nickolai Zeldovich.
\newblock Private web search with tiptoe.
\newblock In {\em Proceedings of the 29th Symposium on Operating Systems Principles}, SOSP '23, page 396–416, New York, NY, USA, 2023. Association for Computing Machinery.

\bibitem{DBLP:conf/sp/Servan-Schreiber22}
Sacha Servan{-}Schreiber, Simon Langowski, and Srinivas Devadas.
\newblock Private approximate nearest neighbor search with sublinear communication.
\newblock In {\em 43rd {IEEE} Symposium on Security and Privacy, {SP} 2022, San Francisco, CA, USA, May 22-26, 2022}, pages 911--929. {IEEE}, 2022.
\newblock \url{https://doi.org/10.1109/SP46214.2022.9833702}.

\bibitem{DBLP:conf/ndss/WuL23}
Zhiqiang Wu and Rui Li.
\newblock {OBI:} a multi-path oblivious {RAM} for forward-and-backward-secure searchable encryption.
\newblock In {\em 30th Annual Network and Distributed System Security Symposium, {NDSS} 2023, San Diego, California, USA, February 27 - March 3, 2023}. The Internet Society, 2023.
\newblock \url{https://www.ndss-symposium.org/ndss-paper/obi-a-multi-path-oblivious-ram-for-forward-and-backward-secure-searchable-encryption/}.

\bibitem{DBLP:conf/osdi/DautermanFLPS20}
Emma Dauterman, Eric Feng, Ellen Luo, Raluca~Ada Popa, and Ion Stoica.
\newblock {DORY:} an encrypted search system with distributed trust.
\newblock In {\em 14th {USENIX} Symposium on Operating Systems Design and Implementation, {OSDI} 2020, Virtual Event, November 4-6, 2020}, pages 1101--1119. {USENIX} Association, 2020.
\newblock \url{https://www.usenix.org/conference/osdi20/presentation/dauterman-dory}.

\bibitem{DBLP:conf/sosp/AhmadSAAG21}
Ishtiyaque Ahmad, Laboni Sarker, Divyakant Agrawal, Amr~El Abbadi, and Trinabh Gupta.
\newblock Coeus: {A} system for oblivious document ranking and retrieval.
\newblock In Robbert van Renesse and Nickolai Zeldovich, editors, {\em {SOSP} '21: {ACM} {SIGOPS} 28th Symposium on Operating Systems Principles, Virtual Event / Koblenz, Germany, October 26-29, 2021}, pages 672--690. {ACM}, 2021.
\newblock \url{https://doi.org/10.1145/3477132.3483586}.

\bibitem{DBLP:conf/sp/MishraPCCP18}
Pratyush Mishra, Rishabh Poddar, Jerry Chen, Alessandro Chiesa, and Raluca~Ada Popa.
\newblock Oblix: An efficient oblivious search index.
\newblock In {\em 2018 {IEEE} Symposium on Security and Privacy, {SP} 2018, Proceedings, 21-23 May 2018, San Francisco, California, {USA}}, pages 279--296. {IEEE} Computer Society, 2018.
\newblock \url{https://doi.org/10.1109/SP.2018.00045}.

\bibitem{bittau2017prochlo}
Andrea Bittau, {\'U}lfar Erlingsson, Petros Maniatis, Ilya Mironov, Ananth Raghunathan, David Lie, Mitch Rudominer, Ushasree Kode, Julien Tinnes, and Bernhard Seefeld.
\newblock Prochlo: Strong privacy for analytics in the crowd.
\newblock In {\em Proceedings of the 26th symposium on operating systems principles}, pages 441--459, 2017.

\bibitem{DBLP:journals/popets/AgarwalHKM19}
Archita Agarwal, Maurice Herlihy, Seny Kamara, and Tarik Moataz.
\newblock Encrypted databases for differential privacy.
\newblock {\em Proc. Priv. Enhancing Technol.}, 2019(3):170--190, 2019.
\newblock \url{https://doi.org/10.2478/popets-2019-0042}.

\bibitem{roth2019honeycrisp}
Edo Roth, Daniel Noble, Brett~Hemenway Falk, and Andreas Haeberlen.
\newblock Honeycrisp: large-scale differentially private aggregation without a trusted core.
\newblock In {\em Proceedings of the 27th ACM Symposium on Operating Systems Principles}, pages 196--210, 2019.

\bibitem{DBLP:conf/uss/0030CDPRR20}
Hao Chen, Ilaria Chillotti, Yihe Dong, Oxana Poburinnaya, Ilya~P. Razenshteyn, and M.~Sadegh Riazi.
\newblock {SANNS:} scaling up secure approximate k-nearest neighbors search.
\newblock In Srdjan Capkun and Franziska Roesner, editors, {\em 29th {USENIX} Security Symposium, {USENIX} Security 2020, August 12-14, 2020}, pages 2111--2128. {USENIX} Association, 2020.
\newblock \url{https://www.usenix.org/conference/usenixsecurity20/presentation/chen-hao}.

\bibitem{DBLP:conf/sigmod/ChowdhuryW0MJ20}
Amrita~Roy Chowdhury, Chenghong Wang, Xi~He, Ashwin Machanavajjhala, and Somesh Jha.
\newblock Crypt$\epsilon$: Crypto-assisted differential privacy on untrusted servers.
\newblock In David Maier, Rachel Pottinger, AnHai Doan, Wang{-}Chiew Tan, Abdussalam Alawini, and Hung~Q. Ngo, editors, {\em Proceedings of the 2020 International Conference on Management of Data, {SIGMOD} Conference 2020, online conference [Portland, OR, USA], June 14-19, 2020}, pages 603--619. {ACM}, 2020.

\bibitem{DBLP:conf/sp/DautermanRPS22}
Emma Dauterman, Mayank Rathee, Raluca~Ada Popa, and Ion Stoica.
\newblock Waldo: {A} private time-series database from function secret sharing.
\newblock In {\em 43rd {IEEE} Symposium on Security and Privacy, {SP} 2022, San Francisco, CA, USA, May 22-26, 2022}, pages 2450--2468. {IEEE}, 2022.
\newblock \url{https://doi.org/10.1109/SP46214.2022.9833611}.

\bibitem{hers}
Joshua~J. Engelsma, Anil~K. Jain, and Vishnu~Naresh Boddeti.
\newblock {HERS:} homomorphically encrypted representation search.
\newblock {\em {IEEE} Trans. Biom. Behav. Identity Sci.}, 4(3):349--360, 2022.
\newblock \url{https://doi.org/10.1109/TBIOM.2021.3139866}.

\bibitem{DBLP:journals/coling/Xiong16}
Deyi Xiong.
\newblock Semantic similarity from natural language and ontology analysis.
\newblock {\em Comput. Linguistics}, 42(4):829--831, 2016.

\bibitem{dwork_algorithmic_2013}
Cynthia Dwork and Aaron Roth.
\newblock The algorithmic foundations of differential privacy.
\newblock 9(3):211--407.

\bibitem{smith_svt}
Adam Smith.
\newblock The sparse vector technique.
\newblock 11 2017.
\newblock \url{https://adaptivedataanalysis.com/wp-content/uploads/2017/11/lect12.pdf}.

\bibitem{DBLP:conf/stoc/DworkNRRV09}
Cynthia Dwork, Moni Naor, Omer Reingold, Guy~N. Rothblum, and Salil~P. Vadhan.
\newblock On the complexity of differentially private data release: efficient algorithms and hardness results.
\newblock In Michael Mitzenmacher, editor, {\em Proceedings of the 41st Annual {ACM} Symposium on Theory of Computing, {STOC} 2009, Bethesda, MD, USA, May 31 - June 2, 2009}, pages 381--390. {ACM}, 2009.
\newblock \url{https://doi.org/10.1145/1536414.1536467}.

\bibitem{DBLP:conf/eurocrypt/DworkKMMN06}
Cynthia Dwork, Krishnaram Kenthapadi, Frank McSherry, Ilya Mironov, and Moni Naor.
\newblock Our data, ourselves: Privacy via distributed noise generation.
\newblock In Serge Vaudenay, editor, {\em Advances in Cryptology - {EUROCRYPT} 2006, 25th Annual International Conference on the Theory and Applications of Cryptographic Techniques, St. Petersburg, Russia, May 28 - June 1, 2006, Proceedings}, volume 4004 of {\em Lecture Notes in Computer Science}, pages 486--503. Springer, 2006.
\newblock \url{https://doi.org/10.1007/11761679\_29}.

\bibitem{Kenthapadi_2013}
Krishnaram Kenthapadi, Aleksandra Korolova, Ilya Mironov, and Nina Mishra.
\newblock Privacy via the johnson-lindenstrauss transform.
\newblock {\em Journal of Privacy and Confidentiality}, 5(1), aug 2013.

\bibitem{lindenstrauss1984extensions}
W~Johnson~J Lindenstrauss.
\newblock Extensions of lipschitz maps into a hilbert space.
\newblock {\em Contemp. Math}, 26(189-206):2, 1984.

\bibitem{evans2018pragmatic}
David Evans, Vladimir Kolesnikov, Mike Rosulek, et~al.
\newblock A pragmatic introduction to secure multi-party computation.
\newblock {\em Foundations and Trends{\textregistered} in Privacy and Security}, 2(2-3):70--246, 2018.

\bibitem{kalogeropoulos2023unraveling}
Antonis Kalogeropoulos and Patr{\'\i}cia Rossini.
\newblock Unraveling whatsapp group dynamics to understand the threat of misinformation in messaging apps.
\newblock {\em New Media \& Society}, page 14614448231199247, 2023.

\bibitem{tan2021tracking}
Edina~YQ Tan, Russell~RE Wee, Young~Ern Saw, Kylie~JQ Heng, Joseph~WE Chin, Eddie~MW Tong, and Jean~CJ Liu.
\newblock Tracking private whatsapp discourse about covid-19 in singapore: longitudinal infodemiology study.
\newblock {\em Journal of Medical Internet Research}, 23(12):e34218, 2021.

\bibitem{DBLP:conf/sp/UngerDBFPG015}
Nik Unger, Sergej Dechand, Joseph Bonneau, Sascha Fahl, Henning Perl, Ian Goldberg, and Matthew Smith.
\newblock Sok: Secure messaging.
\newblock In {\em 2015 {IEEE} Symposium on Security and Privacy, {SP} 2015, San Jose, CA, USA, May 17-21, 2015}, pages 232--249. {IEEE} Computer Society, 2015.

\bibitem{DBLP:conf/acns/VatandasGIK20}
Nihal Vatandas, Rosario Gennaro, Bertrand Ithurburn, and Hugo Krawczyk.
\newblock On the cryptographic deniability of the signal protocol.
\newblock In Mauro Conti, Jianying Zhou, Emiliano Casalicchio, and Angelo Spognardi, editors, {\em Applied Cryptography and Network Security - 18th International Conference, {ACNS} 2020, Rome, Italy, October 19-22, 2020, Proceedings, Part {II}}, volume 12147 of {\em Lecture Notes in Computer Science}, pages 188--209. Springer, 2020.

\bibitem{DBLP:journals/corr/abs-2408-07614}
Kareem Amin, Alex Kulesza, and Sergei Vassilvitskii.
\newblock Practical considerations for differential privacy.
\newblock {\em CoRR}, abs/2408.07614, 2024.

\bibitem{doi:10.1177/2050157920958442}
Antonis Kalogeropoulos.
\newblock Who shares news on mobile messaging applications, why and in what ways? a cross-national analysis.
\newblock {\em Mobile Media \& Communication}, 9(2):336--352, 2021.

\bibitem{grinberg2019fake}
Nir Grinberg, Kenneth Joseph, Lisa Friedland, Briony Swire-Thompson, and David Lazer.
\newblock Fake news on twitter during the 2016 us presidential election.
\newblock {\em Science}, 363(6425):374--378, 2019.

\bibitem{dwork_calibrating_nodate}
Cynthia Dwork, Frank {McSherry}, Kobbi Nissim, and Adam Smith.
\newblock Calibrating noise to sensitivity in private data analysis.

\bibitem{fbsecretconversations}
Facebook.
\newblock Messenger secret conversations technical whitepaper, 5 2017.
\newblock https://about.fb.com/wp-content/uploads/2016/07/messenger-secret-conversations-technical-whitepaper.pdf.

\bibitem{DBLP:conf/chi/FreedPMLRD18}
Diana Freed, Jackeline Palmer, Diana~Elizabeth Minchala, Karen Levy, Thomas Ristenpart, and Nicola Dell.
\newblock "a stalker's paradise": How intimate partner abusers exploit technology.
\newblock In Regan~L. Mandryk, Mark Hancock, Mark Perry, and Anna~L. Cox, editors, {\em Proceedings of the 2018 {CHI} Conference on Human Factors in Computing Systems, {CHI} 2018, Montreal, QC, Canada, April 21-26, 2018}, page 667. {ACM}, 2018.

\bibitem{kasiviswanathan_ldp_2008}
Shiva~Prasad Kasiviswanathan, Homin~K. Lee, Kobbi Nissim, Sofya Raskhodnikova, and Adam Smith.
\newblock What can we learn privately?
\newblock In {\em 2008 49th Annual IEEE Symposium on Foundations of Computer Science}, pages 531--540, 2008.

\bibitem{casacuberta2022widespread}
S{\'\i}lvia Casacuberta, Michael Shoemate, Salil Vadhan, and Connor Wagaman.
\newblock Widespread underestimation of sensitivity in differentially private libraries and how to fix it.
\newblock In {\em Proceedings of the 2022 ACM SIGSAC Conference on Computer and Communications Security}, pages 471--484, 2022.

\bibitem{outsideLookingIn}
Seny Kamara, Mallory Knodel, Emma Llans{\'o}, Greg Nojeim, Lucy Qin, Dhanaraj Thakur, and Caitlin Vogus.
\newblock Outside looking in: Approaches to content moderation in end-to-end encrypted systems.
\newblock {\em arXiv preprint arXiv:2202.04617}, 2022.

\bibitem{scheffler2023sok}
Sarah Scheffler and Jonathan Mayer.
\newblock Sok: Content moderation for end-to-end encryption.
\newblock {\em Proceedings on Privacy Enhancing Technologies}, 2:403--429, 2023.

\bibitem{DBLP:conf/tcc/BonehSW11}
Dan Boneh, Amit Sahai, and Brent Waters.
\newblock Functional encryption: Definitions and challenges.
\newblock In Yuval Ishai, editor, {\em Theory of Cryptography - 8th Theory of Cryptography Conference, {TCC} 2011, Providence, RI, USA, March 28-30, 2011. Proceedings}, volume 6597 of {\em Lecture Notes in Computer Science}, pages 253--273. Springer, 2011.

\bibitem{DBLP:journals/tetc/HanLKMH24}
Kyung~Hyun Han, Wai{-}Kong Lee, Angshuman Karmakar, Jose Maria~Bermudo Mera, and Seong~Oun Hwang.
\newblock cufe: High performance privacy preserving support vector machine with inner-product functional encryption.
\newblock {\em {IEEE} Trans. Emerg. Top. Comput.}, 12(1):328--343, 2024.

\bibitem{DBLP:conf/pkc/AbdallaBCP15}
Michel Abdalla, Florian Bourse, Angelo~De Caro, and David Pointcheval.
\newblock Simple functional encryption schemes for inner products.
\newblock In Jonathan Katz, editor, {\em Public-Key Cryptography - {PKC} 2015 - 18th {IACR} International Conference on Practice and Theory in Public-Key Cryptography, Gaithersburg, MD, USA, March 30 - April 1, 2015, Proceedings}, volume 9020 of {\em Lecture Notes in Computer Science}, pages 733--751. Springer, 2015.

\bibitem{DBLP:journals/amco/MasciaSV23}
Carla Mascia, Massimiliano Sala, and Irene Villa.
\newblock A survey on functional encryption.
\newblock {\em Adv. Math. Commun.}, 17(5):1251--1289, 2023.

\bibitem{DBLP:conf/crypto/AgrawalLS16}
Shweta Agrawal, Beno{\^{\i}}t Libert, and Damien Stehl{\'{e}}.
\newblock Fully secure functional encryption for inner products, from standard assumptions.
\newblock In Matthew Robshaw and Jonathan Katz, editors, {\em Advances in Cryptology - {CRYPTO} 2016 - 36th Annual International Cryptology Conference, Santa Barbara, CA, USA, August 14-18, 2016, Proceedings, Part {III}}, volume 9816 of {\em Lecture Notes in Computer Science}, pages 333--362. Springer, 2016.

\bibitem{DBLP:conf/pkc/MeraKMS22}
Jose Maria~Bermudo Mera, Angshuman Karmakar, Tilen Marc, and Azam Soleimanian.
\newblock Efficient lattice-based inner-product functional encryption.
\newblock In Goichiro Hanaoka, Junji Shikata, and Yohei Watanabe, editors, {\em Public-Key Cryptography - {PKC} 2022 - 25th {IACR} International Conference on Practice and Theory of Public-Key Cryptography, Virtual Event, March 8-11, 2022, Proceedings, Part {II}}, volume 13178 of {\em Lecture Notes in Computer Science}, pages 163--193. Springer, 2022.

\bibitem{DBLP:journals/iacr/FanV12}
Junfeng Fan and Frederik Vercauteren.
\newblock Somewhat practical fully homomorphic encryption.
\newblock {\em {IACR} Cryptol. ePrint Arch.}, page 144, 2012.

\bibitem{DBLP:conf/crypto/Brakerski12}
Zvika Brakerski.
\newblock Fully homomorphic encryption without modulus switching from classical gapsvp.
\newblock In Reihaneh Safavi{-}Naini and Ran Canetti, editors, {\em Advances in Cryptology - {CRYPTO} 2012 - 32nd Annual Cryptology Conference, Santa Barbara, CA, USA, August 19-23, 2012. Proceedings}, volume 7417 of {\em Lecture Notes in Computer Science}, pages 868--886. Springer, 2012.

\bibitem{DBLP:conf/crypto/BrakerskiV11}
Zvika Brakerski and Vinod Vaikuntanathan.
\newblock Fully homomorphic encryption from ring-lwe and security for key dependent messages.
\newblock In Phillip Rogaway, editor, {\em Advances in Cryptology - {CRYPTO} 2011 - 31st Annual Cryptology Conference, Santa Barbara, CA, USA, August 14-18, 2011. Proceedings}, volume 6841 of {\em Lecture Notes in Computer Science}, pages 505--524. Springer, 2011.

\bibitem{viand_verifiable_2023}
Christian Knabenhans, Alexander Viand, Antonio Merino{-}Gallardo, and Anwar Hithnawi.
\newblock vfhe: Verifiable fully homomorphic encryption.
\newblock In Fl{\'{a}}vio Bergamaschi, Anamaria Costache, and Kurt Rohloff, editors, {\em Proceedings of the 12th Workshop on Encrypted Computing {\&} Applied Homomorphic Cryptography, Salt Lake City, UT, USA, October 14-18, 2024}, pages 11--22. {ACM}, 2024.

\bibitem{OpenFHE}
Ahmad~Al Badawi, Jack Bates, Flavio Bergamaschi, David~Bruce Cousins, Saroja Erabelli, Nicholas Genise, Shai Halevi, Hamish Hunt, Andrey Kim, Yongwoo Lee, Zeyu Liu, Daniele Micciancio, Ian Quah, Yuriy Polyakov, Saraswathy R.V., Kurt Rohloff, Jonathan Saylor, Dmitriy Suponitsky, Matthew Triplett, Vinod Vaikuntanathan, and Vincent Zucca.
\newblock Openfhe: Open-source fully homomorphic encryption library.
\newblock Cryptology ePrint Archive, Paper 2022/915, 2022.
\newblock \url{https://eprint.iacr.org/2022/915}.

\end{thebibliography}
\appendix
\label{sec:appendix}

\section{Ethical Considerations}
\label{sec:ethics}

Beyond the privacy preserving methods of \SecretVectorSearch and the importance of direct engagement with data donors, we emphasize the importance of informed ``opt-in'' consent, privacy, and security of analysis of end-to-end encrypted (E2EE) content. Although our purposes are journalistic, technical approaches to measuring trends in E2EE content could be used for a variety of other purposes, from moderation to advertisement to surveillance \cite{outsideLookingIn,scheffler2023sok}.

All components of the data donation model underwent multiple iterations of development to ensure that the storage protocol, communication methods, and consent model satisfied the needs of advocacy groups and on-the-ground data donors. The system was approved via an institutional IRB process and was found to be of minimal risk given our methods and domain of interest.  
Additionally, when preparing to engage with actual data donors, we engaged with advocacy groups in geographical regions of interest to obtain feedback on the user on-boarding system and consent protocol. Our objective was to ensure users had full understanding of potential use cases for their donated data, and what giving consent means. 

A version of this system used only on public WhatsApp groups is currently deployed and in use by journalists. The sample data used in this work was collected from these groups. 

\section{Table of variables}
\Cref{tab:variables} lists the variables used in this work, for convenience.

\begin{table}[!t]
\small
\centering
\def\arraystretch{1}
\caption{Notation used throughout this work.}
\begin{tabular}{p{1cm}p{6.5cm}}%
\toprule
Term & Description \\
\midrule
$\EpsilonFineThreshold$ & Epsilon budget for fine-grained threshold queries  \\
$\EpsilonFineCount$ & Epsilon budget for fine-grained count queries  \\
$\EpsilonCoarse$ & Epsilon budget to build perturbed dataset used in coarse-grained queries \\ 
$\DeltaCoarse$ & Delta budget to build perturbed dataset used in coarse-grained queries \\
$\epsilon$  & Overall epsilon budget \\
$\delta$ & Overall delta budget \\
$\JLParam$ & Distortion induced by the JL transform
\\
\midrule
$\EmbDimInitial$ & Dimension of BERT semantic vector embedding \\
$\EmbDimFinal$ & Dimension of semantic vector post-JL transform (see \Cref{sec:projection}) \\
$\Projection$ & Projection matrix for dimensionality reduction \\ 
$\Delta$ & Gaussian noise matrix for perturbing coarse-grained data \\ 
\midrule
$m$ & Plaintext message \\
$x'$ & $\EmbDimInitial$-length vector which is an embedding of a message $m$.  The vectors together form a database $X'$ \\
$x$ & $\EmbDimFinal$-length vector belonging to a database of message vectors $X = X'P$ \\
$\tilde{x}$ & Perturbed $\EmbDimFinal$-length vector belonging to a database of perturbed message vectors $\tilde{X} = X'P + \Delta$.  Each element of $\Delta$ was sampled randomly from $\mathcal{N}(0, \sigma_\Delta^2)$ where $\sigma_\Delta^2$ is defined in \Cref{eqn:noise-matrix-sigma}. \\
\midrule
$q$ & Query vector (For $q'$ a BERT embedding vector of a plaintext query, $qP$ is the query vector) \\
$a$ & match radius (matches are the set of $x$ for which $\Vert x - q \Vert < a$ (for fine-grained queries, replace $x$ with $\tilde{x}$ for coarse-grained queries) \\
$t$ & threshold (for threshold queries) \\
\bottomrule
\end{tabular}
\label{tab:terminology}
\label{tab:variables}
\end{table}

\section{Additional design considerations}
\label{app:additional-design}

In this section we discuss design considerations that were abridged in \Cref{sec:design-decisions}.

\subsection{Multi-party computation compared with other kinds of cryptography.}
\label{sec:why-mpc}

As described in \Cref{sec:approaching-e2ee} and \Cref{sec:local-and-central}, we identified that one core aspect of \Synopsis must be privacy against the ``service provider'' (data stewards).
For generic private computation, the three main paradigms would be functional encryption (FE), homomorphic encryption (HE), or multi-party computation (MPC).  

Recall that we discussed in \Cref{sec:related-work} that other more targeted paradigms, including searchable encryption, private set intersection, private information retrieval, and secure sketches, did not align with our simultaneous needs to (1) have the system remain private against the querier and server simultaneously, and (2) enable \emph{both prospective and retrospective queries}---since retrospective journalistic analysis requires looking for information that we did not know we wanted at the time the donors input their message into the system, some level of generic computation is required.

Early in design, we considered \emph{functional encryption} (FE; \cite{DBLP:conf/tcc/BonehSW11}) as a candidate. 
FE is defined around a function $f$ of the key $k$ and message $m$; informally the goal of FE is that an encryption $\textsf{FEnc}_f(m, k)$ can only be decrypted to some function $f_k(m)$ rather than decrypting to $m$ itself.
Functional encryption is mostly a theoretical construct with few applied deployments, but several constructions for inner product functionality could be used in applied settings (e.g. \cite{DBLP:journals/tetc/HanLKMH24,DBLP:conf/pkc/AbdallaBCP15, DBLP:journals/amco/MasciaSV23,DBLP:conf/crypto/AgrawalLS16,DBLP:conf/pkc/MeraKMS22}).  A candidate FE-based construction of \Synopsis would have used inner products decrypting to the function of the inner product $(m \cdot k)$ to determine whether a donated message $m$ matched a ``query'' $k$, known as IPFE.

However, we discarded this as an option relatively quickly because of three factors.
First, although IPFE could calculate matches for individual vectors, it did not easily contain functionality to return \emph{counts or thresholds of matches} -- the cryptographic layer would have returned \emph{which} messages were matching, which we wished to avoid.
Second, in FE, one party (usually holding a ``master secret key'') must generate the keys used to do the function decryption.  In general, that party has the ability to create any new key---leading to a decryption of arbitrary functions in the class.  Therefore, we still had to use some kind of non-collusion assumption (assuming that key-holding party would not collude with the servers) --- an assumption that essentially means we are already doing MPC, but a less efficient version.
Third, FE is \emph{significantly} slower than other options; the most efficient existing scheme we found \cite{DBLP:journals/tetc/HanLKMH24} required about 18ms on CPU or 1ms on GPU \emph{to run one decryption of a single element}, meaning that running a match for a single vector would have taken about 10 seconds on CPU or 0.5s on GPU, which would lead to unacceptable query times once the dataset grew into the thousands.

We also considered using (fully) homomorphic encryption (HE;  \cite{DBLP:journals/iacr/FanV12,DBLP:conf/crypto/Brakerski12,DBLP:conf/crypto/BrakerskiV11}), which allows addition and multiplication operations to occur over ciphertexts without needing decryption keys.  HE is typically used in use cases where a querying weak-computation client encrypts data and sends it to a strong-computation server, which performs the addition and multiplication to get a resulting ciphertext, which is returned to the client for decryption.

We also discarded this as an option for three similar factors.
First, because the party who needed to obtain decrypted results (the journalists) was not the party who knew the private information (the data donors), we still needed a non-collusion assumption somewhere in the system, once again meaning we are already doing MPC.
Second, detecting misbehavior by the computing party is surprisingly challenging in HE and often requires re-performing hefty computations to verify or additional heavy zero-knowledge proofs \cite{viand_verifiable_2023}, as opposed to MPC in which malicious security is relatively cheaper.
Third, HE is also significantly slower than MPC.  We tested a version of \Synopsis that used OpenFHE \cite{OpenFHE} for comparison's sake.  It performed better than FE (about 1.26ms per element of a vector, so about 650ms to match an 512-length vector) but still performed significantly worse than the MPC-based implementation we describe in \Cref{sec:implementation} (which matched a comparably-sized vector in just under 1ms total).

Since the other options ultimately boiled down to the same non-collusion threat model as MPC, and were significantly slower, we settled on a (malicious secure) MPC-based implementation using MP-SPDZ \cite{mp-spdz}.

\subsection{On using malicious security despite one organization's control of all code}
\label{sec:malicious-security}
We remark that, from a full computer security perspective, the data stewards can be considered semi-honest code providers.  Since they write the code that collects the data, there is some level of trust that goes beyond what mere security threat models can capture.  (The code is open source, but data donors are not generally expected to review it.)  , we use malicious security for these parties in an MPC sense including a privacy guarantee against the servers.  This is primarily because malicious MPC security provides some party in the system with a signal if something goes wrong---making malicious security practically useful in this scenario even though \emph{from a pure cryptography perspective}, if all MPC server subsidiaries of the \ServerController worked together, they could maliciously update the code to regain access to future raw text messages.

\subsection{Allowing data stewards to view journalists' queries} 
\label{sec:query-privacy}
At present, we choose to make queries known to the server (i.e., public in the MPC computation). We make this choice because the server owners and clients, in our setting, have a real-world relationship and already communicate about in-progress investigations. Treating the query as public also makes the MPC computation more efficient (by performing scalar multiplications rather than secret-secret multiplications; see~\Cref{sec:general_impl} for additional detail). If queries with confidentiality against the servers are needed, our framework can easily accommodate this by treating the journalist's query as a secret input to the MPC, at the cost of higher query runtime.

\section{Synopsis Parties}
The MPC parties in \Synopsis are shown in \Cref{fig:parties} for convenience.
\begin{figure}[t!]
\begin{tikzpicture}

\def\xServers{2}
\def\wServer{0.6}
\def\hServerSeg{0.3}
\def\dServers{1.7}
\def\yServerCenter{0}
\def\ysServers{{-\dServers+\yServerCenter,\yServerCenter,\dServers+\yServerCenter}}

\def\dPhones{1.1}
\def\xPhones{-1}
\def\yPhonesCenter{-0.5}
\def\ysPhones{{\yPhonesCenter-2*\dPhones,\yPhonesCenter-\dPhones, \yPhonesCenter,\yPhonesCenter+\dPhones,\yPhonesCenter+2*\dPhones}}

\def\xaArrowPS{\xPhones+0.5}
\def\xbArrowPS{\xServers-1.3*\wServer}
\def\dyArrowP{\dPhones/1.1/2}
\def\yasArrowPS{{\yPhonesCenter-2*\dPhones+\dyArrowP,\yPhonesCenter-\dPhones+\dyArrowP, \yPhonesCenter+\dyArrowP, \yPhonesCenter+\dPhones+\dyArrowP, \yPhonesCenter+2*\dPhones+\dyArrowP}}
\def\ybsArrowPS{\ysServers}

\def\xJournos{5}
\def\yJournosCenter{0}
\def\dJournos{1.5}
\def\ysJournos{{\yJournosCenter-1.5*\dJournos, \yJournosCenter-0.5*\dJournos, \yJournosCenter+0.5*\dJournos, \yJournosCenter+1.5*\dJournos}}
\def\rJournos{0.5}

\def\yNames{\ysPhones[4]+1.4}

\def\yArrowLabels{\ysPhones[0]-0.4}

\def\xaArrowJS{\xJournos-0.5}
\def\xbArrowJS{\xServers+1.3*\wServer}
\def\dyArrowJ{\dJournos/1.1/2}
\def\yasArrowJS{{\yJournosCenter-1.5*\dJournos, \yJournosCenter-0.5*\dJournos, \yJournosCenter+0.5*\dJournos, \yJournosCenter+1.5*\dJournos}}
\def\ybsArrowJS{\ysServers}

\def\xaLine{\xCloudLimited-3}
\def\xbLine{\xJourno-1}
\def\yLine{\yServerCenter}


\node[database, database radius=\wServer cm, database segment height=\hServerSeg cm] at (\xServers,\ysServers[0]) {};
\node[database, database radius=\wServer cm, database segment height=\hServerSeg cm] at (\xServers,\ysServers[1]) {};
\node[database, database radius=\wServer cm, database segment height=\hServerSeg cm] at (\xServers,\ysServers[2]) {};

\node[smartphone] at (\xPhones,\ysPhones[0]) {};
\node[smartphone] at (\xPhones,\ysPhones[1]) {};
\node[smartphone] at (\xPhones,\ysPhones[2]) {};
\node[smartphone] at (\xPhones,\ysPhones[3]) {};
\node[smartphone] at (\xPhones,\ysPhones[4]) {};

\foreach \a in {0, 1, ..., 4} {
    \foreach \b in {0, 1, 2} {
        \draw[-{Latex[length=1mm]}] (\xaArrowPS, \yasArrowPS[\a]) -- (\xbArrowPS, \ybsArrowPS[\b]+0.1*\a-0.2);
    }
}

\foreach \a in {0, 1, ..., 3} {
    \node[alice, mirrored, minimum size=\rJournos cm] at (\xJournos, \ysJournos[\a]) {};

    \foreach \b in {0, 1, 2} {
        \draw[{Latex[length=1mm]}-{Latex[length=1mm]}] (\xaArrowJS, \yasArrowJS[\a]+0.1*\b-0.2) -- (\xbArrowJS, \ybsArrowJS[\b]+0.1*\a-0.05);
    }
}

\node[] at (\xPhones,\yNames) {Data Donors};
\node[align=center] at (\xServers,\yNames) {MPC Servers\\(controlled by\\Data Stewards)};
\node[align=center] at (\xJournos,\yNames) {Queriers/\\Journalists};

\node[align=center] at (0.5*\xPhones+0.5*\xServers, \yArrowLabels) {Input\\messages};
\node[] at (0.5*\xServers+0.5*\xJournos, \yArrowLabels) {Queries};

\end{tikzpicture}
\caption{Parties in \Synopsis}
\label{fig:parties}
\end{figure}

\section{Privacy budget analysis}
Budget spend on an increasing number of unique queries to the \textit{same database} in the FC and CC query regimes. A one-time ``charge'' of $\EpsilonCountCoarse = 2$ is incurred in the CC regime, in order to generate a perturbed dataset; thereafter, all queries are ``free.'' Each unique FC query incurs an additional charge of $\EpsilonCountFine = 0.6.$

\usepgfplotslibrary{fillbetween}

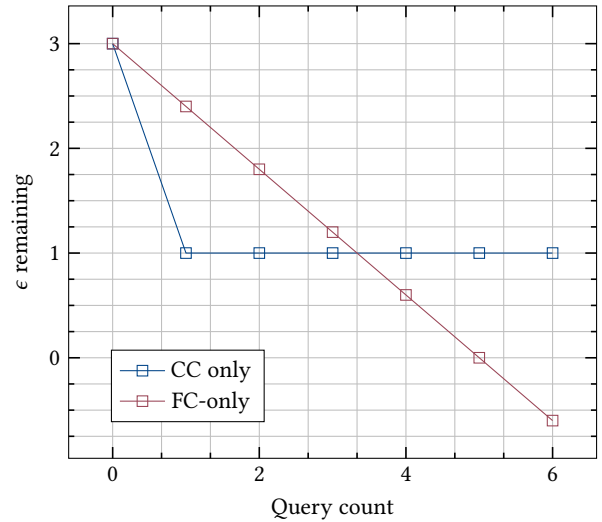
\begin{figure}[t!]
    \centering
        \begin{tikzpicture}
        	\begin{axis}[
        		xlabel= Query count,
        		ylabel=$\epsilon$ remaining,
                    max space between ticks=40,
                    minor x tick num=2,
                    minor y tick num=3,
                    major tick length=0.15cm,
                    minor tick length=0.075cm,
                    tick style={semithick,color=black}, 
                    legend style={at={(0.08,0.16)},anchor=west},
                    height = 7.6 cm,
                    width = 8.6 cm,
                    grid = both,
                    ]

                \addplot+ [ color=\ColorGraphCC,mark=square,
                    name path = a
                ]
                    coordinates {
                    (0, 3) 
                    (1, 1)
                    (2, 1)
                    (3, 1)
                    (4, 1)
                    (5, 1)
                    (6, 1)
        	};
                \addlegendentry{CC only};

                \addplot+ [ color=\ColorGraphFC,mark=square,
                    name path = b
                ]
                    coordinates { 
        	    (0, 3) 
                    (1, 2.4)
                    (2, 1.8)
                    (3, 1.2)
                    (4, 0.6)
                    (5, 0)
                    (6, -0.6)
                };
                \addlegendentry{FC-only};

            \end{axis}

        \end{tikzpicture}
    \caption{Budget spend comparison for FC and CC querying, for at least query issued to a static database of message embeddings. We assume a starting budget of $\epsilon = 3,$ CC queries with $\EpsilonCoarse = 2$, and FC queries with $\EpsilonFineCount = 0.6$ }
    
\end{figure}

\section{\Synopsis Usage Example}
In this section we walk through an example investigation workflow using \Synopsis.
We describe a real-world use case inspired by a news story published by journalist collaborators \cite{jaswal_bjp_2024}.  Note that here the DP privacy ``budget'' has been changed to a more granular number of ``credits.''

\begin{enumerate}
    \item \textbf{On-the-ground conversations [June-November 2022].} Journalist collaborators on the ground in India and the United States negotiate data donation contracts with local community leaders in Ayodhya. 
    Widespread BJP-sponsored messaging has already been noted across public and private WhatsApp groups. 

    \item \textbf{User onboarding [January 2023].} Members of the U.S. team onboard volunteer donors with devices configured to run a prototype version of their donor WhatsApp client. Users can selectively grant this prototype app access to groups to which their account belongs via OAuth. This includes donations from public \textit{and} private groups. 

    \item \textbf{Onboarding journalists [May 2023].}
    A ``privacy budget'' is assigned to the querying user/journalist's account. Information about querying costs, budgeting, and accuracy is relayed to the journalist. The journalist estimates that the time until completion will be approximately 12 months. \textbf{Budget: 150.}

    \item \textbf{Data intake [Ongoing].} The server begins to receive message data in a stream, as messages are sent. 
    As described in \Cref{sec:architecture}, the donor client preprocesses each message $m$ by converting it to a semantic embedding vector $x$, running a noised projection on it to get $\tilde{x}$, dimension reducing both, splitting them into shares, and sending them to the \censor{WhatsAppWatch}-run MPC servers.    

    \item \textbf{Exploratory analysis [July 2023].} As a trial run, the journalists conduct a preliminary analysis of a noisy version of message data collected from June to July 2023. They identify several keywords and topics to monitor throughout the next year, including ``BJP,'' ``Modi,'' ``Ram Temple,'' and ``bulldozers''\footnote{Mentions of bulldozers and other heavy-duty earthmoving equipment often accompany strongly anti-Muslim sentiment related to the construction of the Ram Temple.}. This one-time noising process costs 20 credits. \textbf{Budget: 130.}

    \item \textbf{Creating and monitoring event alerts [September-December 2023].} Because discussions of the BJP and Modi are fairly commonplace in the groups they're monitoring, journalists choose to receive high-count alerts for days when a combination of ``bulldozers'' and ``Ram Temple'' are frequently observed in message data. These alerts only incur a charge when high-count events are detected. In the first four months of monitoring, three high-count events trigger notifications in December. Each notification incurs a charge of 2 credits, for a total of 6 credits spent. \textbf{Budget:~124.}

    \item \textbf{Creating and monitoring real-time trend alerts [January 2024].} Activity in local Ayodhya WhatsApp groups increases in the weeks preceding the inauguration of the Ram Temple. Journalists elect to receive daily trend reports, which display overall hit counts in real time, instead of event alerts. Trend reports cost about 2 credits per day, for a total of 62 credits charged for a month of monitoring. \textbf{Budget: 62.} 

    \item \textbf{Identifying messaging trends and post-analysis [February-March 2024].} Journalists use data collected during these months to home in on a period of one week in mid-January when messaging about the Ram Temple also included conversations about possible offline gatherings. 

    \item \textbf{Creating a noisy dataset for further analysis and public release [May 2024].} As interviews with on-the-ground election monitors and data stewards continue, journalists finish drafting a story about political messaging in WhatsApp groups. In their post hoc analysis, they also identify possible instances of pro- and anti- candidate speech, events, and topics of discussion in the weeks immediately following Narendra Modi's victory. They elect to make a one-time payment to create a noisy version of their dataset to release alongside their published news story. This dataset will be noisier than the version created in July 2023 for exploratory analysis; the cost is 10 credits. \textbf{Budget: 52.}

    \item \textbf{Retrospective analysis and upkeep.} The team opts to shut off data collection at the conclusion of the study. They decide to use their remaining 52 credits for targeted analysis of collected data. 
    
\end{enumerate}

\section{Full ideal functionality}
The ideal functionality shown in \Cref{fig:ideal-func} ommitted some details required for the MPC simulation to hold, but unnecessary to understand the functionality.  The full ideal functionality is shown in \Cref{alg:full-ideal-func}

\begin{figure}[h]
\fbox{%
\begin{minipage}{0.97\linewidth}
\footnotesize
\textbf{Full ideal functionality for \Synopsis including simulation for MPC security proof.  MPC details, which are the differences from the shortened version in \Cref{alg:ideal-func}, are shown in \textcolor{blue}{blue}.)} \\
\textbf{Parties}: \emph{Donors} submit input messages, \emph{Servers} controlled by the data stewards store shares of the fine-grain and coarse-grain dataset of messages, \emph{Journalists} make queries and receive outputs of queries. \\
\textbf{Setup}:
\begin{itemize}[left=0.1cm]
\item Set maximum per-epoch fine-grain budget $\EpsilonFineTotal$.
\item Set coarse-grain budget $\EpsilonCoarseTotal$, generate projection matrix $P$ based on $\EpsilonCoarseTotal$ as described in \Cref{alg:kkmm}.
\item When reaching a new epoch $\Epoch$, initialize empty coarse database $\DBCoarseEpoch$, and initialize empty fine database $\DBFineEpoch$ with budget $\EpsilonFEpoch = \EpsilonFineTotal$.
\end{itemize}
\textbf{Upon receiving a new donated message $m$ during epoch $\Epoch$ from a donor}:
\begin{itemize}[left=0.1cm]
\item Get message embedding $x$ from $m$.  Store $x$ in the fine-grain database $\DBFineEpoch$.
\item Compute perturbed embedding $\tilde{x} = x + r$ (for multi-variate Gaussian $r$) as described in \Cref{alg:storage}. Store $\tilde{x}$ in the coarse-grain database $\DBCoarseEpoch$.
\item \textcolor{blue}{Create $\NParties$ pairs of random elements (simulated shares of $x$ and $\tilde{x}$) and send these to each MPC server for associate with epoch $\Epoch$.}
\end{itemize}
\textbf{Upon receiving a coarse-grain count query (CC)} (query point=$q$, radius=$a$) for epoch $\Epoch$ from a journalist:
\begin{itemize}[left=0.1cm]
\item Let $c$ be the count of entries $\tilde{x}$ within $\DBCoarseEpoch$ that are within distance $a$ of $q$.
\item \textcolor{blue}{Send the query ($q, a$) to the Servers and wait for their response. If all respond consistently with their simulated $\tilde{x}$ shares, send $c$ to the journalist. If a corrupt server responded inconsistently, instead send $\textsf{Error}$.}
\end{itemize}
\textbf{Upon receiving a coarse-grain threshold query (CT)} (query point=$q$, radius=$a$, threshold $t$) for epoch $\Epoch$ from a journalist:
\begin{itemize}[left=0.1cm]
\item Let $h$ be the threshold result, i.e., if $\DBCoarseEpoch$ contains at least $t$ entries $\tilde{x}$ that are within distance $a$ of $q$, $h=1$, else $h=0$.
\item \textcolor{blue}{Send the query ($q, a, t$) to the Servers and wait for their response. If all respond consistently with their simulated $\tilde{x}$ shares, send $h$ to the journalist. If a corrupt server responded inconsistently, instead send $\textsf{Error}$.}
\end{itemize}
\textbf{Upon receiving a fine-grain count query (FC)} (query point=$q$, radius=$a$) for epoch $\Epoch$ with budget $b$ from a journalist:
\begin{itemize}[left=0.1cm]
\item Check the current remaining budget $\EpsilonFEpoch$.  If it is less than $b$, return $\bot$ and exit.  Else, reduce $\EpsilonFEpoch$ by $b$.
\item Let $c$ be the count of the number of entries $x$ within $\DBFineEpoch$ that are within distance $a$ of $q$.
\item Roll Laplace noise $s$ based on $b$ as described in \Cref{alg:query-fine-count}.
\item \textcolor{blue}{Send the query ($q, a, b$) to the Servers and wait for their response. If all respond consistently with their simulated $\tilde{x}$ shares, send $c+s$ (the noised count) to the journalist. If a corrupt server responded inconsistently, instead send $\textsf{Error}$.}
\item If $\EpsilonFEpoch$ is now 0 (all budget is spent for this epoch), permanently delete $\DBFineEpoch$ \textcolor{blue}{and instruct each Server to delete the corresponding simulated $x$ values}.  (All future queries for $\DBFineEpoch$ will be refused.)
\end{itemize}
\textbf{Upon receiving a fine-grain threshold query (FT)} (query point=$q$, radius=$a$, threshold=$t$) for epoch $\Epoch$ with budget $b$ from a journalist:
\begin{itemize}[left=0.1cm]
\item Check the current remaining budget $\EpsilonFEpoch$.  If it is less than $b$, return $\bot$ and exit.
\item Let $c$ be the count of the number of entries $x$ within $\DBFineEpoch$ that are within distance $a$ of $q$.
\item Roll Laplace noise $u$ and $v$ based on $b$ as described in \Cref{alg:query-fine-threshold}.
\item Let $h$ be 1 if $c + v \ge t + u$, 0 otherwise (i.e. the noised threshold was met/exceeded by the noised count).
\item \textcolor{blue}{Send the query ($q, a, t, b$) to the Servers and wait for their response. If all respond consistently with their simulated $x$ shares, send $h$ to the journalist. If a corrupt server responded inconsistently, instead send $\textsf{Error}$.}
\item If $h=1$ and there was no $\textsf{Error}$, lower $\EpsilonFEpoch$ by $b$ and inform the servers that the threshold was exceeded.
\item If $\EpsilonFEpoch$ is now 0 (all budget is spent for this epoch), permanently delete $\DBFineEpoch$ \textcolor{blue}{and instruct all servers to delete all fine-grain shares of each $x$ in that epoch}. (All future queries for $\DBFineEpoch$ will be refused.)
\end{itemize}%
\end{minipage}%
} 
\caption{Ideal Functionality for \Synopsis}
\label{alg:ideal-func-full}
\label{alg:full-ideal-func}
\end{figure}

\section{Additional Algorithms from Section 5} 
\label{app:coarse-figures}
\label{app:extra-algs}
These algorithms contain the full details of the algorithms omitted in \Cref{sec:architecture}.  In particular \Cref{alg:storage} shows the process of storing the queries (which involves storing some extra information to make MPC queries easier), and \Cref{alg:cc,alg:ct} show the two coarse grain query mechanisms ommitted in \Cref{sec:coarse-grain}.
\begin{figure}
\fbox{\begin{minipage}{0.97\linewidth}
\footnotesize
\textbf{Algorithm}: Storage and pre-query processing of messages: 
\\
\textbf{Summary}: Donors submit messages that are stored as shares between the MPC servers.
Messages are row vectors. 
There is some redundancy in the data stored: storing $x^2$ in addition to $x$ improves the efficiency of the MPC versions of the DP mechanisms shown in Figs.\ \ref{alg:fc}, \ref{alg:ft}, \ref{alg:cc}, and \ref{alg:ct}, and storing the noised $\tilde{x}$ in addition to $x$ enables coarse-grained queries.
\\
\textbf{Parameters}: 
Initial embedding length $\EmbDimInitial$.
Final embedding length $\EmbDimFinal$.
Public random $\EmbDimInitial \times \EmbDimFinal$ projection matrix $\Projection$ chosen as described in \Cref{alg:noise-generation}.
Privacy parameters $(\EpsilonCoarse, \DeltaCoarse)$.
Scaling factor $\ScalingFactor$.
\\
\textbf{Inputs}: 
\begin{itemize}[left=0.1cm]
    \item\textbf{Donors}: Each donor $d$ inputs plaintext messages $m_1, m_2, ..., m_{n_d}$  
    \item\textbf{Servers}: No input 
\end{itemize}
\textbf{Outputs}: 
\begin{itemize}[left=0.1cm]
    \item\textbf{Donors}: No output
    \item\textbf{Servers}:  After the protocol, for each message, the servers hold secret shares of:
    \begin{itemize}[left=0.1cm]
        \item $x$ ($k$-length embedding vector for a message)
        \item $x^2$ (elementwise square of $x$)
        \item $\tilde{x}$ ($x$ plus Gaussian noise)
        \item $\tilde{x}^2$ (elementwise square of $\tilde{x}$)
    \end{itemize}
\end{itemize}
\textbf{Algorithm}:
\begin{enumerate}[left=0.1cm]
    \item Donor processing (build $x$ and $x^2$ from messages, shares of perturbation $r$, perturbed points $\tilde{x} = x+r$ and $\tilde{x}^2$ from $x$ and $r$).
    \begin{itemize}
        \item Each donor computes BERT message embedding $x'$ of each message $m$.
        \item Each donor computes $x = \ScalingFactor x'P$.  ($\ScalingFactor$ is a scalar parameter based on the embedding model that ensures each entry of $x$ is betwen $[-1,1]$)
        \item Each donor computes $x^2$, the element-wise square of $x$
        \item Each donor samples $r$ as an $\EmbDimFinal$-length vector where each element is sampled from $\mathcal{N}(0, \sigma_\Delta)$ where $\sigma_\Delta$ is set by $\epsilon_J$, $\delta_J$, and $\EmbDimFinal$ as in \Cref{eqn:noise-matrix-sigma}.
        \item Each donor sets $\tilde{x} = b'(x + r)$ where $b'$ is a scalar parameter to normalize the vector to length 1.
        \item Each donor sends secret shares $[x]$, $[x^2]$, $[\tilde{x}]$, $[\tilde{x}^2]$ to the servers.
    \end{itemize}
    \item Server processes messages (in preparation to receive future queries)
        Server stores $[x]$ and $[x^2]$ to use in future fine-grained queries, and stores $[\tilde{x}]$ and $[\tilde{x}^2]$ to use in future coarse-grained queries.
\end{enumerate}
\textbf{Privacy budget burned}: $\EpsilonCoarse$ (and $\DeltaCoarse$) burned to create the perturbed $\tilde{x}$ dataset
\end{minipage}}
\caption{Algorithm: Storage and pre-query processing}
\label{alg:storage}
\label{fig:storage}
\end{figure}
\begin{figure}[h]
\fbox{\begin{minipage}{0.97\linewidth}
\footnotesize
\textbf{Algorithm}: Coarse-grained count query
\\
\textbf{Summary}: Within a particular epoch, querier learns a count of perturbed elements matching query $q$ (i.e. the number of $\tilde{x}$ points within L2 distance $a$ of $q$).
\\
\textbf{DP Mechanism}: PrivateProjection mechanism (\cite{Kenthapadi_2013} Alg.\ 1, a variant of the Gaussian mechanism \cite{DBLP:conf/eurocrypt/DworkKMMN06})
\\
\textbf{Parameters}: $\sigma_\Delta^2$ (a parameter setting the magnitude of added Gaussian noise described in \Cref{alg:kkmm})
\\
\textbf{Inputs}: 
\begin{itemize}[left=0.1cm]
    \item\textbf{Servers}: Stored $[\tilde{x}]$ and $[\tilde{x}^2]$ shares
    \item\textbf{Querier}: Query point $q$, radius $a'$ (both public)
\end{itemize}
\textbf{Outputs}: 
\begin{itemize}[left=0.1cm]
    \item\textbf{Servers}: No output
    \item\textbf{Querier}: The count of perturbed points $\tilde{x}$ such that $\Dist(x, q) < a$
\end{itemize}
\textbf{Algorithm}:
\begin{enumerate}[left=0.1cm]
    \item Follow steps \ref{alg:fc-start2}-%
    \ref{alg:fc-d} of \QueryFineCount, except replace the $[x]$ and $[x^2]$ shares with $[\tilde{x}]$ and $[\tilde{x}^2]$ shares respectively.  At this point, for each point $\tilde{x}$, the servers currently hold shares of $d$, the squared distance between $\tilde{x}$ and $q$. \label{alg:cc-start}
    \item \label{alg:cc-cprime} Follow steps \ref{alg:fc-b}-
    \ref{alg:fc-cprime} of \QueryFineCount.  At this point the servers have $[\tilde{c}']$, the count of elements $\tilde{x}$ that are a match with $q$.
    \item Servers return $[\tilde{c}']$  to the querier. 
    \item Querier reconstructs and outputs $\tilde{c}'$
\end{enumerate}
\textbf{Privacy budget burned}: 0.  (Earlier, $(\EpsilonCoarse, \DeltaCoarse)$ was spent in the creation of the $\tilde{x}$ database; all queries to the $\tilde{x}$ burn no privacy budget.)
\end{minipage}}
\caption{Algorithm: \QueryCoarseCount}
\label{alg:cc}
\label{alg:query-coarse-count}
\end{figure}
\begin{figure}[h]
\fbox{\begin{minipage}{0.97\linewidth}
\footnotesize
\textbf{Algorithm}: Coarse-grained threshold query
\\
\textbf{Summary}: Within a particular epoch, and for given threshold $t$, querier learns a bit representing whether count of perturbed messages matching a query is above or below $t$
\\
\textbf{DP Mechanism}: PrivateProjection mechanism (\cite{Kenthapadi_2013} Alg.\ 1, a variant of the Gaussian mechanism \cite{DBLP:conf/eurocrypt/DworkKMMN06})
\\
\textbf{Parameters}: $\sigma_\Delta^2$ (a parameter setting the magnitude of added Gaussian noise described in \Cref{alg:kkmm})
\\
\textbf{Inputs}: 
\begin{itemize}[left=0.1cm]
    \item\textbf{Servers}: Stored $[\tilde{x}]$ and $[\tilde{x}^2]$ shares
    \item\textbf{Querier}: Query point $q$, radius $a$, threshold $t$ (all public)
\end{itemize}
\textbf{Outputs}: 
\begin{itemize}[left=0.1cm]
    \item\textbf{Servers}: No output
    \item\textbf{Querier}: 1 if the count of perturbed points $\tilde{x}$ such that $\Dist(x, q) < a$ is above $t$, 0 otherwise
\end{itemize}
\textbf{Algorithm}:
\begin{enumerate}[left=0.1cm]
    \item Follow steps \ref{alg:cc-start}-\ref{alg:cc-cprime} of \QueryCoarseCount.
    At this point the servers have $[\tilde{c}']$, the count of elements $\tilde{x}$ that are a match with $q$.
    \item Servers compute $[\tilde{\tau}] = [\tilde{c}' > t]$, shares of a Boolean value which is 1 if the count $c'$ is above the threshold $t$, , 0 otherwise. \label{alg:ct-thresh}
    \item Servers send $[\tilde{\tau}]$ to querier.
    \item Querier reconstructs and outputs $\tilde{\tau}$.
\end{enumerate}
\textbf{Privacy budget burned}: 0 (Earlier, $(\EpsilonCoarse, \DeltaCoarse)$ was spent in the creation of the $\tilde{x}$ database; all queries to the $\tilde{x}$ burn no privacy budget.)
\end{minipage}}
\caption{Algorithm: \QueryCoarseThreshold}
\label{alg:ct}
\label{alg:query-coarse-threshold}
\end{figure}

\section{Consent dialogue for public group data collection}
\begin{figure}
    \centering
    \includegraphics[width=0.9\linewidth]{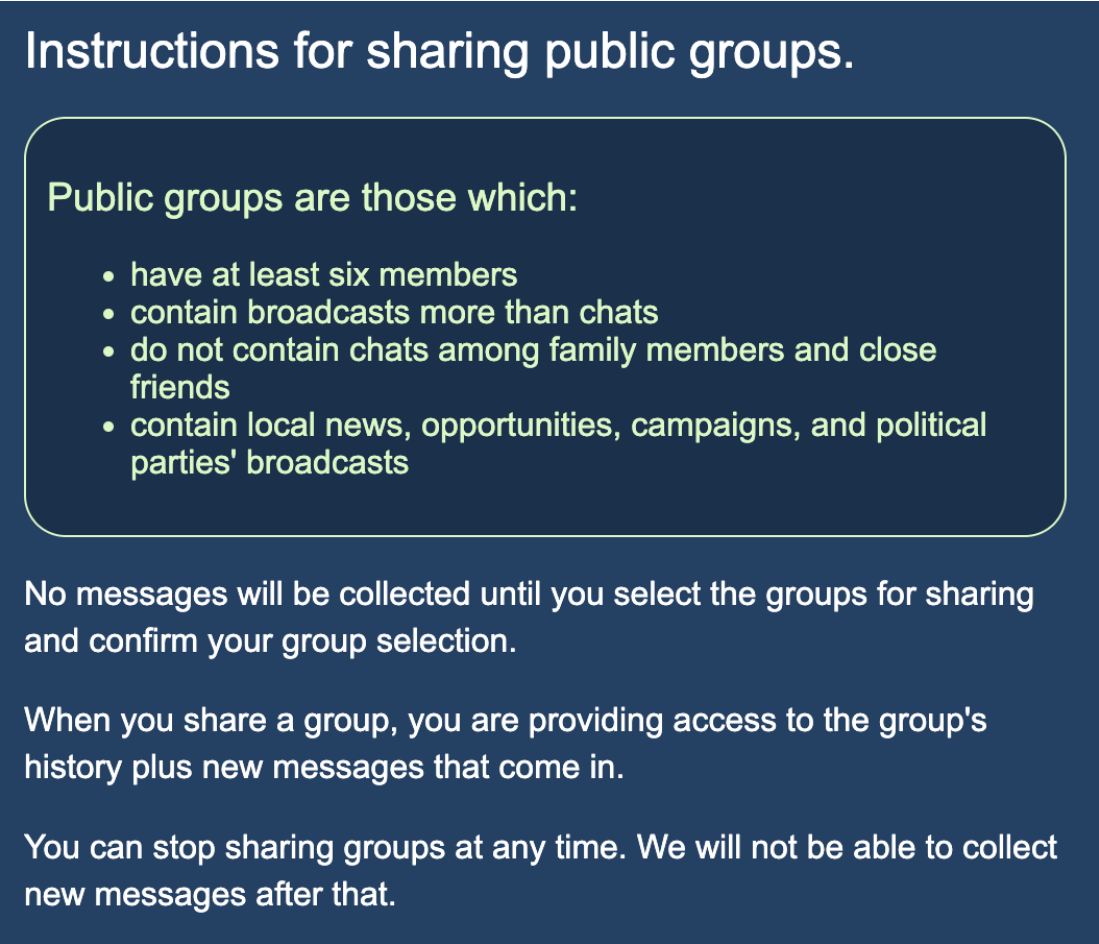}
    \caption{Consent dialogue for public group data collection.}
    \label{fig:waw_dialogue}
\end{figure}

See \Cref{fig:waw_dialogue} to see one portion of the consent dialogue for public WhatsApp group data collection.

\section{Additional details on cosine distance}
\label{app:cosine-facts}

We use cosine distance (a number between 0 and 2) in place of cosine similarity when it is more convenient to work with distance rather than similarity.

We remark that, when $\Vert a \Vert = \Vert b \Vert = 1$, cosine similarity is simply $(a \cdot b)$, the dot product between $a$ and $b$ and also the cosine of the angle between $a$ and $b$.

We also remark on the relationship between L2 distance and the cosine similarity for vectors $a$, $b$ where $\Vert a \Vert = \Vert b \Vert = 1$:
\begin{align*}
\Vert a - b \Vert^2 &= ((a - b) \cdot (a - b)) \\
&=
(a \cdot a) - (a \cdot b) + (b \cdot b) - (b \cdot a) \\
&= \Vert a \Vert^2 + \Vert b \Vert^2 - 2(a \cdot b) \\
&= 2 - 2(a \cdot b)
\end{align*}

\end{document}